\documentclass[twocolumn,superscriptaddress,showpacs,aps]{revtex4-1}
\usepackage{ulem,bm}
\usepackage{amsmath, upgreek, amssymb, graphicx, paralist, gensymb,epstopdf}
\usepackage[colorlinks, linkcolor=blue, citecolor=blue, urlcolor=blue]{hyperref}
\usepackage{array,epsfig,amsmath,amssymb}
\usepackage{graphicx} 
\usepackage{dsfont}
\usepackage{color}

\def\bra#1{\langle #1|}
\def\ket#1{|#1 \rangle}

\newcommand{\braket}[2]{\left\langle#1\right\vert#2\rangle}

\begin{document}

\title{Fundamental building block for all-optical scalable quantum networks}

\author{Seung-Woo Lee}\email{swleego@gmail.com}
\affiliation{Quantum Universe Center, Korea Institute for Advanced Study, Seoul 02455, Korea}

\author{Timothy C. Ralph}
\affiliation{Centre for Quantum Computation and Communication Technology, School of Mathematics and Physics, University of Queensland, St Lucia, Queensland 4072, Australia}

\author{Hyunseok Jeong}
\affiliation{Department of Physics and Astronomy, Seoul National University, Seoul 08826, Korea}

\begin{abstract}

Major obstacles against efficient long distance quantum communication are photon losses during transmission and the probabilistic nature of Bell measurement causing exponential scaling in time and resource with distance. To overcome these difficulties, while conventional quantum repeaters require matter-based operations with long-lived quantum memories, recent proposals have employed encoded multiple photons in entanglement, providing an alternative way for scalability. In pursuing scalable quantum communications, naturally arising questions are thus whether any ultimate limit exists in all-optical scalability, and whether and how it can be achieved. Motivated by these questions, we derive the fundamental limits of the efficiency and loss-tolerance of the Bell measurement with multiple photons, restricted not by protocols but by the laws of physics, i.e.~linear optics and no-cloning theorem. We then propose a Bell measurement scheme with linear optics, which enables one to reach both the fundamental limits: one by linear optics and the other by the no-cloning theorem. The quantum repeater based on our scheme allows one to achieve fast and efficient quantum communication over arbitrary long distances, outperforming previous all-photonic and matter-based protocols. 
Our work provides a fundamental building block for quantum networks within but toward the ultimate limits of all-optical scalability.

\end{abstract}

\maketitle


\section{Introduction}
Quantum communication offers fundamentally secure data transmissions \cite{BB84,Ekert91} and faithful transfers of quantum states \cite{Teleportation}. These are the key elements in building quantum networks as a backbone for promising quantum information protocols \cite{Nielsenbook} such as quantum cryptography \cite{Gisin2002,Lo2014} and distributed quantum computation \cite{Kimble2008}. Developing reliable and efficient quantum communications from within metropolitan areas to over continental scales has been thus one of the important scientific and technological challenges \cite{Pirandola2015,QiChao2016,Valivarthi2016}. Photons are ideal carriers for quantum communication. However, there have been two major obstacles to scalability in photonic quantum communication. One is {\em `photon loss'} during transmission. The survival rate of traveling photons decays exponentially with distance. Even very high-repetition-rate sources (e.g.~$10$ GHz) yield only very low transmission rates at remote places (about $1.8\times 10^{-10}$ Hz at $1,000$ km via optical fibers, i.e.,~only 1 photon arrives every 175 years). In contrast to classical communications, the quantum state of photons cannot be amplified due to the no-cloning theorem \cite{Wootters1982}. The other is {\em `non-deterministic Bell measurement'} with single photons. The Bell measurement is an essential requirement to extend the communication range of photons by quantum teleportation \cite{Lut99} or entanglement swapping \cite{Zukowski93,Pan98}. However, its success probability with single-photon encoding cannot exceed 50\% with linear optics \cite{Weinfurter94,Calsa2001}. As a result, all-optical approaches to quantum communication have suffered from exponential scaling in time and resources with distance \cite{Jacobs2002,Riedmatten2004}.

To overcome these obstacles, a quantum repeater -- a device to extend the communication range with polynomial scaling -- has been developed \cite{Sangouard11,Briegle98,DLCZ,Dur99,Kok2003,Simon2007,Jiang2009,Munro10,Munro12,Muralidharan14,Muralidharan16,Azuma15,Ewert16,Zwerger16}. It works as a building block at intermediate nodes, constituting an entire communication network. In the standard quantum repeater model \cite{Sangouard11,Briegle98,DLCZ,Dur99,Kok2003,Simon2007}, transmission losses are circumvented through heralded entanglement generation between nodes with the help of long-lived quantum memories. Instead, some recent proposals employ quantum error correction schemes with multiple photons \cite{Jiang2009,Munro10,Munro12,Muralidharan14,Muralidharan16,Azuma15,Ewert16,Zwerger16}. In this approach, encoded multiple photons are transmitted between nodes, and losses (as well as other errors) are corrected in the repeater. Quantum repeater protocols developed in this direction could enhance the performance further without use of long-lived quantum memories \cite{Munro12,Muralidharan14,Muralidharan16,Azuma15,Ewert16}. 

A multi-photon encoding approach hence opens the possibility of all-optical scalability. Both photon loss and the probabilistic nature of Bell measurement can be circumvented to some extent through entanglement of photons. The Knill-Laflamme-Milburn protocol \cite{Knill2001} showed that the failure probability of Bell measurements can be reduced to $1/(N + 1)$ with linear optics and $N$ entangled photons. Advanced Bell measurement schemes using additional entangled photons or alternative encoding strategies have been proposed to reach even further beyond the 50\% limit \cite{Azuma15,Ewert16,Jeong2001,SLee13,Grice2011,Zaidi2013,Ewert2014,SLee15,SLee15a}. All-optical quantum repeaters, categorized also as 3rd generation, have been recently developed \cite{Azuma15,Ewert16} and demonstrated \cite{Yasushi19,Zehng19}: A repeater protocol proposed by Azuma {\em et al.} \cite{Azuma15}, employing photonic cluster-states to overcome probabilistic Bell measurements and an additional code for loss-tolerance \cite{Varnava2006}, could achieve a comparable performance with the fastest matter-based repeater \cite{Munro12}. A proposal by Ewert {\em et al.} \cite{Ewert16} based on the parity state error correction encoding \cite{Ralph2005}, in principle, enables an ultra-fast communication without feedforward assuming instant generations of entangled photons. These repeater protocols based on optical systems provide some advantages as discussed in \cite{Azuma15,Ewert16} as they can be performed by photon sources, linear optical elements, and photon detectors. Since a deterministic conversion between photon and matter qubits is demanding, an all-optical approach with entangled photonic qubits at room temperature may be quite an attractive alternative route towards scalable quantum communications, along with the progress of photon source and detector technologies \cite{Eisaman2011,Aharonovich2016,Senellart2017}. 

Therefore, in pursuing scalable quantum networks, the question that comes to mind is whether any fundamental limits exist in the realization of quantum communication with optical components and many photons. One may also wonder whether and how the limits can be reached (if they exist). In this article, we address these questions. We derive, for the first time, the fundamental limits of all-optical scalability in quantum communication. These limits are determined not by protocols but by the laws of physics, i.e., linear optics and the no-cloning theorem. We then propose a Bell measurement scheme with linear optics and multi-photon encoding, which surpasses all the previous schemes and allows us to reach both the fundamental limits. We finally show that the quantum repeater based on our scheme enables fast and efficient quantum communication over arbitrary long distances, outperforming all the previous quantum repeater protocols. Our work thus provides a fundamental building block for quantum networks towards reaching the ultimate limits of all-optical scalability. The main achievements of our work are described below: 

(i) We first derive the fundamental upper bounds of the efficiency and loss-tolerance of Bell measurement in Section~\ref{sec:Limits}. The 50\% limit of Bell measurement with linear optics \cite{Weinfurter94,Calsa2001} is not true anymore when using multiple photons. We thus prove that the maximum success probability of Bell measurement is $1-2^{-N}$ with linear optics and $N$-photon encoding, as the generalization of the limit for the Bell measurement with single photons ($N=1$) \cite{Calsa2001}. We also show that the loss-tolerance of Bell measurement process (Bell measurement with any error correction scheme) is fundamentally limited by $\eta\eta'>0.5$ due to the no-cloning theorem, when photon loss occurs generally at both qubits with different rates $\eta$ and $\eta'$. This is another but general manifestation (in the context of Bell measurement) of the no-cloning limit shown within some error correction protocols \cite{Varnava2006,Stace2009}. These two limits not only determine the ultimate limit of all-optical scalability in quantum communication but also are valid for any quantum information protocols using the Bell measurement on photons.

(ii) We then propose a Bell measurement scheme with linear optics and multi-photon encoding in a concatenated manner in Section~\ref{sec:CBM}, which will be referred to as {\em concatenated Bell measurement} (CBM). It enables one to discriminate Bell states near-deterministically and loss-tolerantly, outperforming all other proposals \cite{Knill2001,Grice2011,Zaidi2013,Ewert2014,SLee15,SLee15a,Ewert16} with respect to the attained success probability with given number of photons and loss rate. Note that CBM is the first and so far the only Bell measurement saturating both fundamental limits by optimization. The scheme is highly compatible with current optical technologies, as it can be performed by the standard Bell measurement \cite{Calsa2001,Weinfurter94} and feedforward controls.

(iii) We then construct a building block of quantum networks (either for transmitting information along the network or for distributing entanglement across the network) based on CBM in Section~\ref{sec:BBQN}. It does not require long-lived quantum memories, photon-matter interactions, nor complicated circuit operations. The communication protocol is optimized by numerical searches, taking into account errors and losses not only during transmission but also during stationary process such as resource generation and measurement in the repeater. It exhibits exponential superiority over conventional quantum relays \cite{Jacobs2002,Riedmatten2004} and several (at least five or six) order of magnitude better performance than standard quantum repeaters \cite{Sangouard11}. Remarkably, it also outperforms all the recent advanced matter-based \cite{Munro12} and all-optical \cite{Azuma15} protocols; it costs an order of magnitude less ($\sim18\%$) photons to achieve near the best performance of those protocols, or yields almost unit transmission probability with similar cost. 

Finally, we conclude with remarks on the potential impacts of our work in developments towards scalable quantum networks in Section~\ref{sec:con}. The ongoing development of the entangled photon sources \cite{Schwartz2016,Collins2013,Silverstone2015,Pichler2017,Buterakos2017} and quantum technologies with integrated optics \cite{Prevedel2007,Silverstone2014,Najafi2015,Carolan2015,Silverston16,Rudolph2016} are expected to enhance the feasibility and performance of our protocol. We will also discuss future studies and proof-of-principle tests based on our work in Section~\ref{sec:con}.  

\section{Fundamental limits}
\label{sec:Limits}

We start with the derivation of the fundamental limits of all-optical scalability. The Bell measurement plays a key role in building scalable architecture as long as photon sources are prepared. Note that the effects of imperfections and errors that photons undergo propagate before being measured and are emerged in detection events. We can thus evaluate the limits by assessing the performance of Bell measurements in each building block. We address the upper bounds of the success probability and the loss-tolerance of Bell measurement restricted by linear optics and the no-cloning theorem, respectively, as follows.

\subsection{Linear optics}
\label{sec:lolimit}

\begin{figure}
\centering
\includegraphics[width=3.4in]{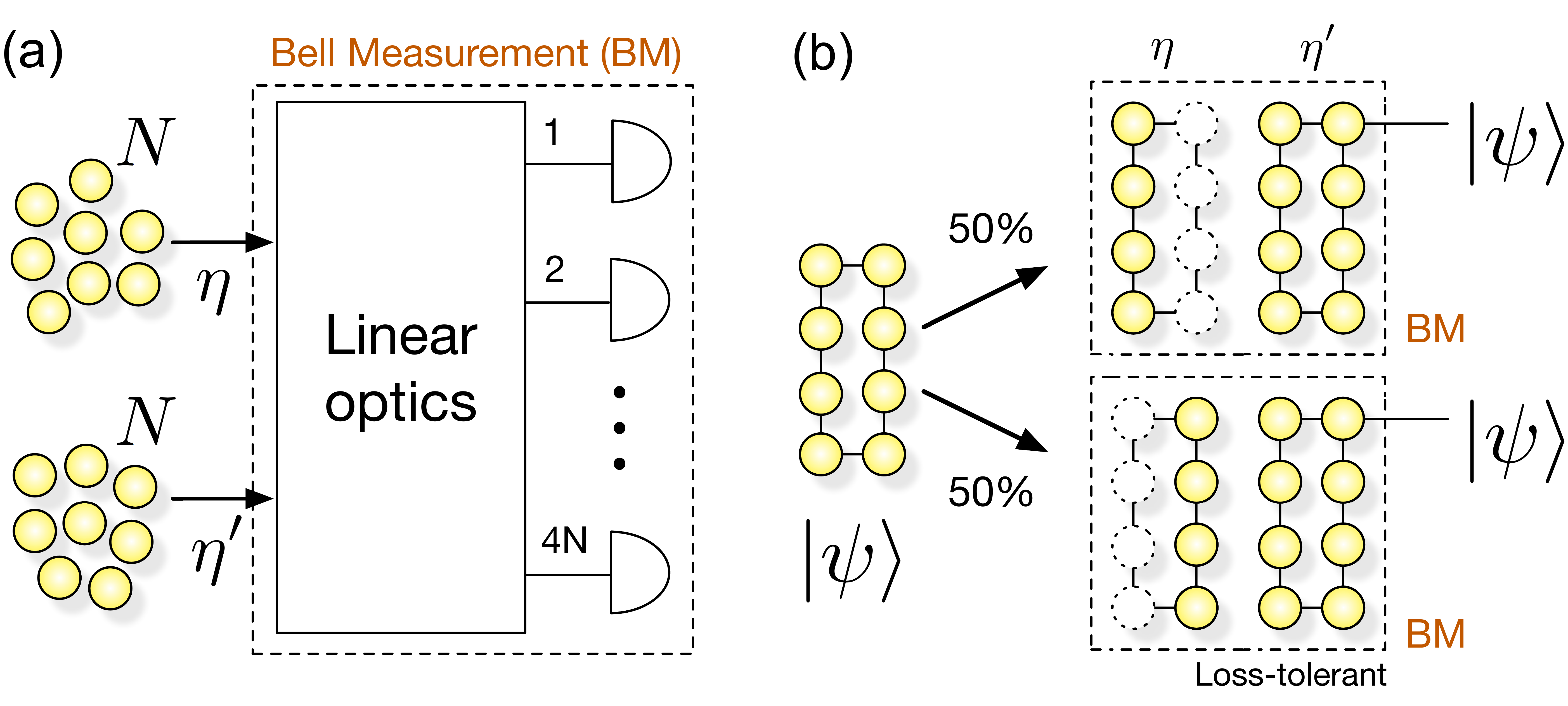}
\caption{Fundamental limits of Bell measurement. (a) General Bell measurement setup with linear optics and $N$-photon encoding. Two qubits (containing total 2N photons) enter $4N$ input modes of a linear optical device in the dual-rail representation, and are detected at $4N$ output modes (by detectors resolving up to two photons). The transmission probabilities of photons in two qubits are $\eta$ and $\eta'$. Then, the maximum success probability of the Bell measurement is obtained as $1-2^{-N}$. (b) If any Bell measurement were able to tolerate 50\% (or more) loss of photons, i.e.,~$\eta\eta'\leq0.5$, it would violate the no-cloning theorem.
}\label{fig:FBounds}
\end{figure}

We consider a general Bell measurement setup illustrated in Fig.~\ref{fig:FBounds}(a). Two qubits, each containing $N$ photons, are prepared equiprobably in a logical Bell state out of four. The logical basis are assumed to be encoded without redundancy such that they are generally written by $\ket{0_L}={\cal U}a^{\dag}_{i_1}\cdots a^{\dag}_{i_N}\ket{0}$ and $\ket{1_L}={\cal U}a^{\dag}_{j_1}\cdots a^{\dag}_{j_N}\ket{0}$ in the dual-rail representation, where $i_1,\cdots,i_N,j_1,\cdots,j_N$ denote the mode numbers given by a permutation of 1 to 2N, and ${\cal U}$ is an arbitrary unitary operation. Total $2N$ photons occupying $4N$ modes enter the linear-optical device and are detected at each output modes. The creation operator of the output modes $\{\hat{c}^{\dag}_k\}$ is then represented by the linear combination of the creation operators of the input modes $\{\hat{a}^{\dag}_i\}$, {\em i.e.} $\hat{c}^{\dag}_k=\sum^{4N}_iU_{ki}a^{\dag}_i$ where $U$ is the unitary matrix for the linear-optical device. The detectors are assumed to resolve up to 2 photons (i.e. 0, 1, and $\geq 2$) to meet the minimum requirement of the standard Bell measurement with single photons \cite{Weinfurter94,Calsa2001}. This is a necessary assumption for the proper generalization of the linear optical limit (see Appendix~\ref{asec:dreso}).

Our aim is to find the maximum success probability of the Bell measurement for arbitrary $N$. For evaluating the success probability of a given Bell measurement setup, we investigate the detection events at single output mode $\hat{c}^{\dag}_k$ and define their corresponding conditional states. In general, if the conditional state yielded from one Bell state is linearly independent to the conditional states from the others, the corresponding detection event allows us to unambiguously discriminate the Bell state. Therefore, the maximum number of linearly independent conditional states of all the possible detection events determines the maximum success probability of the Bell measurement. For example, the maximum success probability of the Bell measurement with single photons ($N=1$) was proved as $1/2$ in this manner in Ref.~\cite{Calsa2001}. 

For the proof of the cases with $N>1$ as detailed in Appendix~\ref{asec:limit}, we can use the following: 

{\bf A1.}~The detection events that more than two photons arrive at any single output mode reduce the success probability of the Bell measurement. This is because such an event cannot be distinguished from the loss of the surplus photon(s) with detectors resolving up to two photons. We can thus restrict the linear-optical map of the setup to a certain class $\{U\}$ by which only less than two photons arrive at each output mode to find the maximum of the success probability of the Bell measurement.

{\bf A2.}~The success probability to discriminate the Bell states defined with the logical basis $\ket{0_L}={\cal U}a^{\dag}_{i_1}\cdots a^{\dag}_{i_N}\ket{0}$ and $\ket{1_L}={\cal U}a^{\dag}_{j_1}\cdots a^{\dag}_{j_N}\ket{0}$ for arbitrary ${\cal U}$ is upper bounded by the maximum success probability to discriminate the Bell states defined with the basis for ${\cal U}=I$.

{\em Proof.}--The Bell states defined with arbitrary ${\cal U}$ can be represented as a superposition of the Bell states with ${\cal U}=I$  (see Appendix~\ref{asec:limit}). So, any conditional state $\ket{\Phi}$ corresponding to a detection event yielded from the input Bell state with arbitrary ${\cal U}$ is in a linear combination of the conditional states $\ket{\phi}$ yielded from the Bell states with ${\cal U}=I$. Since all the element of $\{\ket{\Phi}\}$ is a linear combination of the element of $\{\ket{\phi}\}$, the number of linearly independent $\ket{\Phi}$ is upper bounded with the number of linearly independent $\ket{\phi}$. The probability to unambiguously discriminate the Bell states with any ${\cal U}$ is thus, at best, the maximum success probability to discriminate the Bell states with ${\cal U}=I$.

We can prove $P_s\leq1-2^{-N}$ with arbitrary $N$ as detailed in Appendix~\ref{asec:limit}. This is, to our knowledge, the first proof of the maximum success probability of the Bell measurement with linear optics and multiple photons, and the generalization of the limit $1/2$ proved for $N=1$ in Ref.~\cite{Calsa2001}. The bound might seem to be achieved by multiplexing (i.e.~when any single success out of $N$ trials with single photons is regarded as the overall success), but this is not the case because the encoded Bell state is random in every trial. Notably, it was observed in Ref.~\cite{SLee15} that $1-2^{-N}$ is reachable with photons in the GHZ entanglement, and any ancillary usage of photons yields lower probabilities than this.

\subsection{No-cloning theorem}
\label{sec:NCT}

Let us now derive the fundamental limit of the loss-tolerance of the Bell measurement. The two input qubits are subject to losses respectively with transmission probability $\eta$ and $\eta'$ as illustrated in Fig.~\ref{fig:FBounds}(a). Assume that a quantum error correction performed prior to or during the Bell measurement process. The success probability of the Bell measurement is then given by $P_s(\eta,\eta')$, by which we can prove the limit as follows:

{\bf B1.}~
$P_s(\eta,\eta')=P_s(\eta\eta')$, $\forall$$\eta$ and $\eta'$. 

{\em Proof.}--$P_s(\eta,\eta')$ can be evaluated by summing all the contribution of events at the detectors to identifying the input Bell state. First, any success event is conditioned on the survival of photons from both qubits with probability $\eta\eta'$. Second, some loss events can also contribute to discriminate Bell states by error correction: From the fact that photons from two qubits are indistinguishable, any loss detection event does not tell us which qubit the loss occurs in. This corresponds to the symmetry condition, $P_s(\eta,\eta')=P_s(\eta',\eta)$. Another necessary condition is a failure with null input, $P_s(\eta,0)=0$. Therefore, the loss-tolerance is not individually dependent on the rate of each different loss events, i.e., $\eta(1-\eta')$ or $(1-\eta)\eta'$ or $(1-\eta)(1-\eta')$, but is given with the overall loss rate $1-\eta\eta'$. As a result, $P_s(\eta,\eta')$ is a function of $\eta\eta'$.

{\bf B2.}~The loss-tolerance of Bell measurement is limited by $\eta\eta'>0.5$ due to the no-cloning theorem.

{\em Proof.}--Assume a loss-tolerant Bell measurement which can correct losses up to $P_s(\eta\eta')=1$ for certain $\eta\eta'<1$. We can apply it to quantum teleportation of an unknown qubit $\ket{\psi}$ containing multiple photons. If such a Bell measurement were available for $\eta\eta'\leq0.5$, the teleportation would succeed as long as 50\% photons of $\ket{\psi}$ survive (in the assumption that the channel states are prepared with $\eta'=1$). Then, it would become possible to copy $\ket{\psi}$ deterministically, by dividing the photons of $\ket{\psi}$ into halves to teleport as illustrated in Fig.~\ref{fig:FBounds}(b), which violates the no-cloning theorem.

The loss-tolerance limit $\eta\eta'>0.5$ is fundamental so that any Bell measurement with the help of error correction technique can never exceed. While the 50\% no-cloning limit has been discussed in other contexts \cite{Varnava2006,Stace2009,Muralidharan16}, our result is generally valid for any model containing joint measurements under photon losses.

\section{Concatenated Bell measurement}
\label{sec:CBM}

We here propose a Bell measurement scheme with linear optics in a concatenated manner (referred to as concatenated Bell measurement, CBM). In our approach, the parity state encoding \cite{Ralph2005} is employed with the logical basis $\ket{0_L}=\ket{+^{(m)}}^{\otimes n}$ and $\ket{1_L}=\ket{-^{(m)}}^{\otimes n}$, where $\ket{\pm^{(m)}}=\ket{H}^{\otimes m}\pm\ket{V}^{\otimes m}$ (the coefficient will be omitted unless necessary). Each logical qubit contains $n$ blocks of $m$ (total $N=nm$) photons. Following the decomposition procedure in Ref.~\cite{SLee15}, the logical Bell states, $\ket{\Phi^{\pm}}=\ket{0_L}\ket{0_L}\pm\ket{1_L}\ket{1_L}$ and $\ket{\Psi^{\pm}}=\ket{0_L}\ket{1_L}\pm\ket{1_L}\ket{0_L}$, can be completely decomposed into the block size Bell states, $\ket{\phi^{\pm}_{(m)}}=\ket{+^{(m)}}\ket{+^{(m)}}\pm\ket{-^{(m)}}\ket{-^{(m)}}$ and $\ket{\psi^{\pm}_{(m)}}=\ket{+^{(m)}}\ket{-^{(m)}}\pm\ket{-^{(m)}}\ket{+^{(m)}}$, which are also completely decomposed into the Bell states with photon pair, $\ket{\phi^{\pm}}=\ket{+}\ket{+}\pm\ket{-}\ket{-}$ and $\ket{\psi^{\pm}}=\ket{+}\ket{-}\pm\ket{-}\ket{+}$ (see Appendix~\ref{asec:Decom}). We denote the logical, block size, photon pair Bell states as the 2nd, 1st, 0th level Bell states, respectively. The Bell states in higher levels can be fully characterized by the type and number of lower level Bell states that appear in the decomposition (see Table~\ref{table:BMresult}).

\begin{table}[h]
\caption{\label{table:BMresult} \textbf{Bell states decomposition}}
\centering
\begin{ruledtabular}
\begin{tabular}{l  c  c}
Level & Bell states  & Decomposed into\\[0.5ex]
\hline
2nd  & & {\small even(odd) number of} $\ket{\phi^{-}_{(m)}}$, \\[-2ex]
(logical) & \raisebox{1.5ex}{$\ket{\Phi^{+(-)}}$} &and $\ket{\phi^{+}_{(m)}}$ for others\\
\cline{2-3}
& & even(odd) number of $\ket{\psi^{-}_{(m)}}$,\\[-2ex]
& \raisebox{1.5ex}{$\ket{\Psi^{+(-)}}$} &and $\ket{\psi^{+}_{(m)}}$ for others\\
\hline
1st &  & even number of $\ket{\psi^{+(-)}}$, \\[-2ex]
(block) & \raisebox{1.5ex}{$\ket{\phi^{+(-)}_{(m)}}$} &and $\ket{\phi^{+(-)}}$ for others\\
\cline{2-3}
 & & odd number of $\ket{\psi^{+(-)}}$, \\[-2ex]
 & \raisebox{1.5ex}{$\ket{\psi^{+(-)}_{(m)}}$} &and $\ket{\phi^{+(-)}}$ for others\\[-0.5ex]
\end{tabular}
\end{ruledtabular}
\end{table}

\begin{figure}
\centering
\includegraphics[width=3.4in]{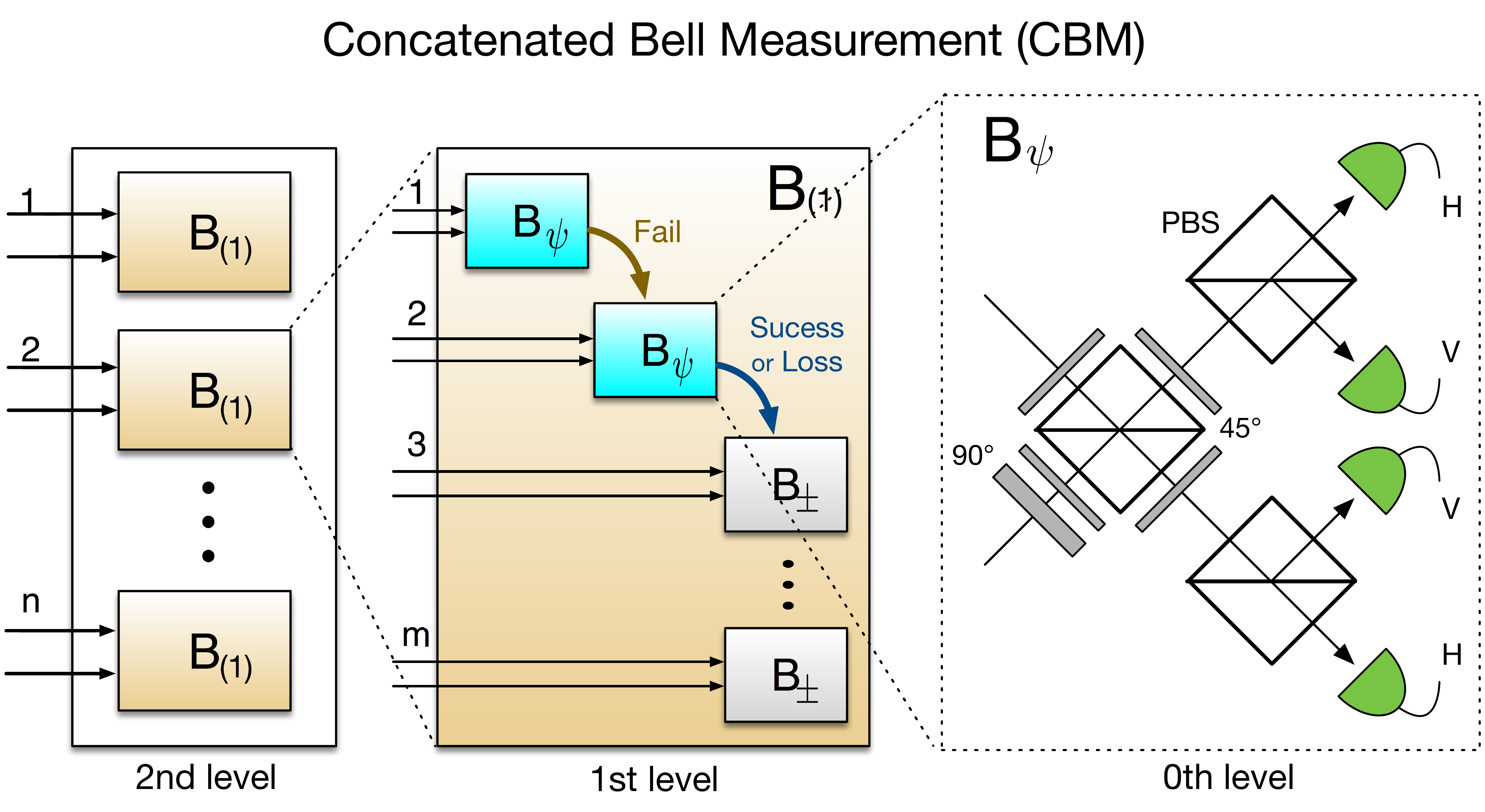}
\includegraphics[width=3.4in]{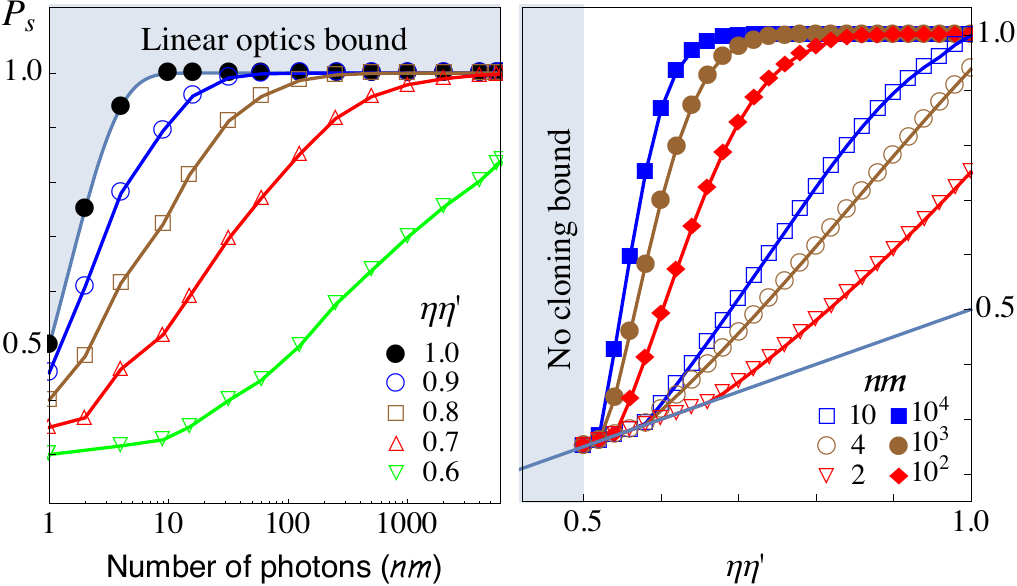}
\caption{Bell measurement scheme in a concatenated manner with linear optics. The (logical) 2nd level Bell measurement ${\rm B}_{(2)}$ is composed of $n$ independent ${\rm B}_{(1)}$ measurements, each of which is performed with $m$-times of 0th level Bell measurements ${\rm B}_{(0)}=\{{\rm B}_{\psi}, {\rm B}_{+}, {\rm B}_{-}\}$ with feedforwards. Three types of ${\rm B}_{(0)}$ are the variations of the standard Bell measurement scheme with linear optics. If we remove all the wave plates at two input modes of the first PBS in ${\rm B}_{\psi}$, it becomes ${\rm B}_{+}$, while if we remove only the two $45^\circ$ wave plates, it becomes ${\rm B}_{-}$.
Bottom: The maximum success probability $P_s$ of optimized CBM is plotted (Left) against the number of photons $N=nm$ in a qubit for different loss rates $\eta\eta'$, (Right) against $\eta\eta'$ for different $nm$. The solid line is the success probability of the Bell measurement on single photons, $\eta\eta'/2$. 
}\label{fig:CBM}
\end{figure}

\subsection{Scheme}

Let us describe the CBM scheme (see Appendix~\ref{sec:detailCBM} for details). This is composed of three concatenated levels as illustrated in Fig.~\ref{fig:CBM}, i.e., 0th level is for photon size, 1st is for block size, and 2nd level is for the logical encoding size:

{\em (0th level)} For the 0th level Bell measurement (referred as ${\rm B}_{(0)}$), we employ the standard scheme of Bell measurement using linear optics elements such as polarizing beam splitter, wave plates and photon detectors \cite{Lut99,Calsa2001}. It enables to unambiguously discriminate two 0th level Bell states out of the four, $\ket{\phi^{\pm}}$ and $\ket{\psi^{\pm}}$. The two identified Bell states can be chosen by changing the wave plates at the input modes. We define three different types ${\rm B}_{(0)}=\{{\rm B}_{\psi}, {\rm B}_{+}, {\rm B}_{-}\}$ that respectively discriminate $\{(\ket{\psi^+},\ket{\psi^-}), (\ket{\phi^+},\ket{\psi^+}), (\ket{\phi^-},\ket{\psi^-})\}$ with the success probability $1/2$ in an ideal case.

{\em (1st level)} In the 1st level, ${\rm B}_{(1)}$, total $m$-times of ${\rm B}_{(0)}$ are performed. First, ${\rm B}_{\psi}$ is applied to arbitrary photon pair (one from the first qubit and the other from the second) repeatedly until it succeeds or detects a loss or consecutively fails $j$-times ($0\leq j\leq m-1$). Then, either ${\rm B}_{+}$ or ${\rm B}_{-}$ is applied on the remaining photon pairs; ${\rm B}_{\pm}$ is selected if ${\rm B}_{\psi}$ succeeded with $\ket{\psi^\pm}$, while arbitrary one is chosen for loss detection or $j$-times failures. Note that $j$ can be selected to optimize the scheme for a given number of photons $nm$ and loss rate $\eta$.  

The result of ${\rm B}_{(1)}$ is determined as: (Success) Full discrimination of the 1st level Bell states is possible unless loss occurs. For example, if any ${\rm B}_{\psi}$ succeeds with $\ket{\psi^{+}}$, subsequently performed all ${\rm B}_{+}$ should yield either $\ket{\phi^{+}}$ or $\ket{\psi^{+}}$. From the Table~\ref{table:BMresult}, if even(odd) number of $\ket{\psi^{+}}$ appear, one can find that the 1st level Bell state is $\ket{\phi^{+}_{(m)}}$($\ket{\psi^{+}_{(m)}}$). (Sign $\pm$ discrimination) As long as any single $\rm B_{\psi}$ succeeds or any $\rm B_{\pm}$ is performed without loss, the $\pm$ sign can be identified. (Failure) $\rm B_{(1)}$ fails when no $\rm B_{\psi}$ succeeds and all ${\rm B}_{\pm}$ detect losses.

We denote the success and failure probability of $\rm B_{(1)}$ respectively as $p_s$ and $p_f$, and thus the probability of sign $\pm$ discrimination only is given by $1-p_f-p_s$.

{\em (2nd level)} The 2nd (logical) level, $\rm B_{(2)}$, is composed of $n$ independent $\rm B_{(1)}$. It is constructed such that loss in any $\rm B_{(1)}$ does not affect the other $\rm B_{(1)}$. The Bell states, $\ket{\Phi^{\pm}}$ and $\ket{\Psi^{\pm}}$, can be unambiguously discriminated as long as any single $\rm B_{(1)}$ succeeds and no $\rm B_{(1)}$ fails, so that the success probability is $P_s=(1-p_f)^n-(1-p_s-p_f)^n$. The result is given as $\ket{\{\Phi,\Psi\}^{(-)^s}}$. Here, the symbol $\Phi$ and $\Psi$ is discriminated by any success of $\rm B_{(1)}$. The sign $(-)^s$ is then identified if $s$ (either even or odd) number of minus($-$) signs appear among all $\rm B_{(1)}$ results. For example, given the results of $\rm B_{(1)}$ as $\{\ket{\phi_{(m)}^{-}},+,-\}$, $\ket{\Phi^{+}}$ can be identified by $\phi$ and even number of minus($-$) signs. 

\subsection{Reaching the fundamental limits} 

If all the photons in two qubits survive ($\eta=\eta'=1$), no ${\rm B}_{(1)}$ would fail (at least sign discrimination is possible), i.e.,~$p_f=0$. It would succeed unless all performed $\rm B_{\psi}$ fail and the subsequent choice, either $\rm B_{+}$ or $\rm B_{-}$, is wrong with probability $1/2$, such that $p_s=1-2^{-j-1}$. The overall success probability is then obtained as $P_s=1-2^{-(j+1)n}$. If we set $j=m-1$, it attains $P_s=1-2^{-N}$ the fundamental upper bound limited by linear optics with $N=nm$. 

For arbitrary $\eta$ and $\eta'$, the success and failure probability of $\rm B_{(1)}$ are obtained as $p_s=(1-2^{-(j+1)})\eta^m\eta'^m$ and $p_f=\sum^{m}_{l=m-j}(\eta\eta'/2)^{m-l}(1-\eta\eta')^l$, respectively (details in Appendix~\ref{asec:sPCBM}). The overall success probability of CBM is then obtained by $P_s(\eta,\eta')=(1-p_f)^n-(1-p_s-p_f)^n$. Note that, as expected from {\bf B1} in Section~\ref{sec:NCT}, the Bell measurement succeeds with the same probability as long as $\eta\eta'$ is the same, i.e.,~$P_s(\eta,\eta')=P_s(\eta\eta')$. The maximum success probabilities are plotted in Fig.~\ref{fig:CBM} by optimization over $\{n,m,j\}$. We can observe that arbitrary high success probabilities up to unit is reachable, as long as $\eta\eta'>0.5$, by increasing $N=nm$ within the linear optics bound. 

Therefore, it turns out that CBM reaches both fundamental limits by linear optics and no-cloning theorem. From a practical point of view, it outperforms all the previous Bell measurement schemes \cite{Ewert16,SLee15}, with respect to the success probability obtained by the same number of photons per qubit with a given loss rate as presented in Appendix~\ref{asec:compBM}. Note that the linear optics bound can be saturated when no photon loss occurs in our scheme. Note also that, in contrast to other schemes consuming more redundant photons under losses \cite{Ewert16,SLee15}, all photons in CBM effectively contribute either for success events or loss-tolerance. 

\subsection{General error correction}

Logical errors (bit or/and sign flips) can be produced due to depolarization and imperfect operations. These emerge as either sign ($+ \leftrightarrow -$) or symbol ($\phi \leftrightarrow \psi$) flip in the result of CBM. Some sign flip errors that occur in any $\rm B_{(0)}$ can be corrected by majority vote in the 1st level, based on the fact that the signs in all $\rm B_{(0)}$ results should be the same within an ideal $\rm B_{(1)}$. Symbol flips (although odd number of symbol flips in $\rm B_{(0)}$ cause a symbol flip in $\rm B_{(1)}$) can be eventually corrected in the 2nd level, from the fact the symbols of all $\rm B_{(1)}$ in any ideal $\rm B_{(2)}$ result should be the same. This is because the parity state encoding is a generalized form of the Shor error correcting code \cite{Shor95}. As a result, logical errors can be reduced in the result of CBM as long as $n,m\geq3$ without any additional process (see Appendix~\ref{asec:LECBM}). We also observed that the effect of dark counts, caused by detector imperfection, can be reduced (and negligible) in CBM as described in Appendix~\ref{asec:dcCBM}.

\section{Building blocks for quantum network}
\label{sec:BBQN}
In this section, we propose a protocol for all-optical scalable quantum communications as the main result of this article. We intend to use CBM for developing building blocks of scalable quantum networks. Our approach is based on the following concepts: a logical photonic qubit in parity state encoding \cite{Ralph2005} is employed as the carrier of information. It enables us to achieve higher transmission rates under losses than single photons. This is, however, limited to a certain distance range depending on the encoding size. To extend the range further, quantum repeaters are needed to be placed at appropriate intervals. In our protocol, we employ CBM. We note that CBM is able to extend both the communication rate and range over direct transmissions with a given number of photons (see Appendix~\ref{asec:extension}). 

\begin{figure}[b]
\centering
\includegraphics[width=2.2in]{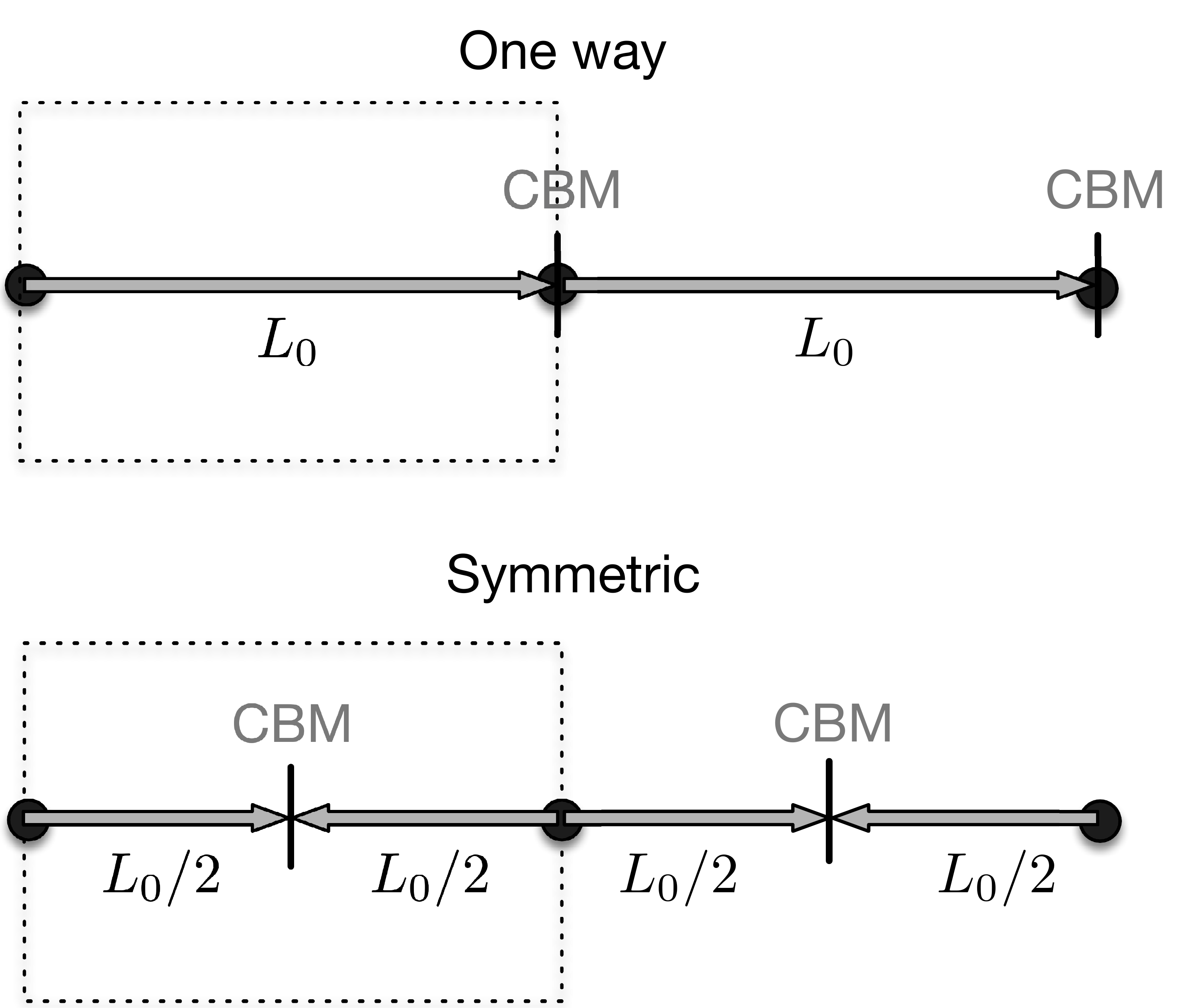}
\caption{Two designs of quantum networks. (One way) a qubit travels $L_0$ between nodes along the network. (Symmetric) two qubits travel $L_0/2$ to meet in the middle where CBM is performed for entanglement distribution across the network.
}\label{fig:2Types}
\end{figure}

In each building block, loss occurs not only during transmission but also during other processes such as resource generation and measurement. We thus consider the effective survival rate of individual photons in whole stationary process as $\eta_0$. In general, a photon survives with probability $\eta_L=\eta_0e^{-L/L_{att}}$ in one cycle of the generation, transmission (over distance $L$), and measurement process, where $L_{\rm att}$ is the attenuation length.

\subsection{Quantum network designs}
\label{sec:twodesigns}

\begin{figure*}
\centering
\includegraphics[width=1\linewidth]{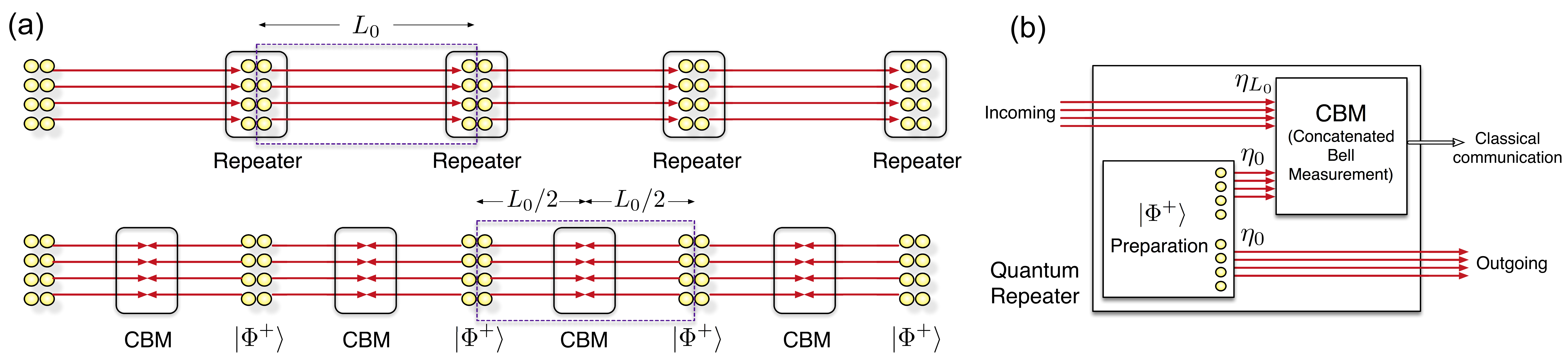}
\caption{Schematic of building blocks for quantum network. (a) Two designs of building blocks for quantum networks: (Top) for one-way communication to transmit quantum information along the network, in which the qubit carrying information travels $L_0$ between repeater nodes. The other qubit is staying in the repeater. Then, CBM is performed on the transmitted and stationary qubits. 
(Bottom) for entanglement distribution between remote places, in which CBM is performed to link the entangled pairs $\ket{\Phi^+}$ from adjacent nodes. Each qubit travels $L_0/2$ to meet in the middle before CBM. 
Note that both designs of building blocks yield the same success probabilities, $P_s(\eta_{L_0},\eta_0)=P_s(\eta_{L_0/2},\eta_{L_0/2})$ and cost the same number of photons on average. (b) A quantum repeater for one-way communication is composed of two parts: the preparation of entangled pair $\ket{\Phi^+}$ and CBM on two qubits (one is received from the previous node and the other from $\ket{\Phi^+}$). The other qubit of $\ket{\Phi^+}$ is transmitted to the next node. The result of CBM is directly sent to Bob via classical communication based on which the transmitted information can be recovered at the final step. Losses during preparation and measurement process also affect the performance. The effective loss rate of photons inside of the repeater is referred as $\eta_0$, estimated with source efficiency $\epsilon_s$, detector efficiency $\epsilon_d$, and the loss during the generation time of $\ket{\Phi^+}$.
}\label{fig:LDQC}
\end{figure*}

Two different designs of quantum networks can be considered:  i) One way: this is for transmitting quantum information along the network as illustrated at the top of Fig.~\ref{fig:2Types}. In each building block of this design, CBM is performed on two qubits; one qubit travels between repeater nodes (say $L_0$), while the other remains stationary in the repeater. Its success probability is given by $P_s(\eta_{L_0},\eta_0)$. ii) Symmetric: this is for the entanglement distribution by entanglement swapping across the network as illustrated at the bottom of Fig.~\ref{fig:2Types}. In each building block, CBM is performed to link the entangled pairs between adjacent nodes. Two qubits from different nodes travel over $L_0/2$ to meet in the middle before CBM, so its success probability is $P_s(\eta_{L_0/2},\eta_{L_0/2})$. The success probabilities of these two designs are exactly the same $P_s(\eta_{L_0},\eta_0)=P_s(\eta_{L_0/2},\eta_{L_0/2})$ as $\eta_{L_0}\eta_0=\eta_{L_0/2}^2=\eta_0^2 e^{-L_0/L_{att}}$. Therefore, any of these two (or their combination) can be chosen as a building block to construct a quantum network depending on the application purpose as illustrated in Fig.~\ref{fig:LDQC}(a).

\subsection{Quantum repeater}
\label{sec:repeater}

Let us describe the details of our protocol with a realistic repeater model. We will focus on the {\em one way} type of communication  for transmitting quantum information from Alice to Bob (but the estimated performance will be the same with the {\em symmetric} type). The total distance $L$ between Alice and Bob is divided into $L_0$ by equally spaced nodes. 

Our repeater model is illustrated in Fig.~\ref{fig:LDQC}(b), which is composed of two parts, one is for the preparation of a logical Bell pair $\ket{\Phi^+}$ and the other is for CBM. Notably, it does not require long-lived quantum memories, photon-matter interactions, nor complicated circuit operations. In a realistic system, the losses and imperfections during the process in the repeater (in addition to the attenuation during transmission), which both qubits experience before CBM, can strongly influence the performance of the repeater. In each repeater, CBM is applied between the incoming qubit and one qubit from $\ket{\Phi^+}$ so that the success probability of each building block is $P_s(\eta_{L_0},\eta_0)$. 

Here, $\eta_0$ is the effective survival rate of photons during the stationary process in the repeater, which can be estimated by 
\begin{equation}
\begin{aligned}
\eta_0(n,m)=\epsilon_s\epsilon_d \exp\Big[-\frac{c(\tau_p(n,m)+\tau)}{L_{\rm att}}\Big],
\end{aligned}
\label{eq:epsilon1}
\end{equation}
where $\epsilon_s$ and $\epsilon_d$ are the source and detector efficiency, respectively, and $\exp[-c(\tau_p+\tau)/L_{\rm att}]$ denotes the rate that individual photon survives during the preparation and measurement process, and $c$ is the speed of light. Here $\tau$ is the time taken for single or two photon measurement with appropriate feedforward, and $\tau_p(n,m)$ is the estimated average time for the preparation of the logical Bell pair $\ket{\Phi^+}$. The rate $\eta_0(n,m)$ is given as a function of the encoding size $(n,m)$ because the more photons are contained in a logical qubit the longer time $\tau_p$ will be taken to generate the state $\ket{\Phi^+}$ as detailed in Appendix~\ref{asec:timetau}.

\subsection{Performance analysis} 

\begin{figure}
\centering
\includegraphics[width=3.5in]{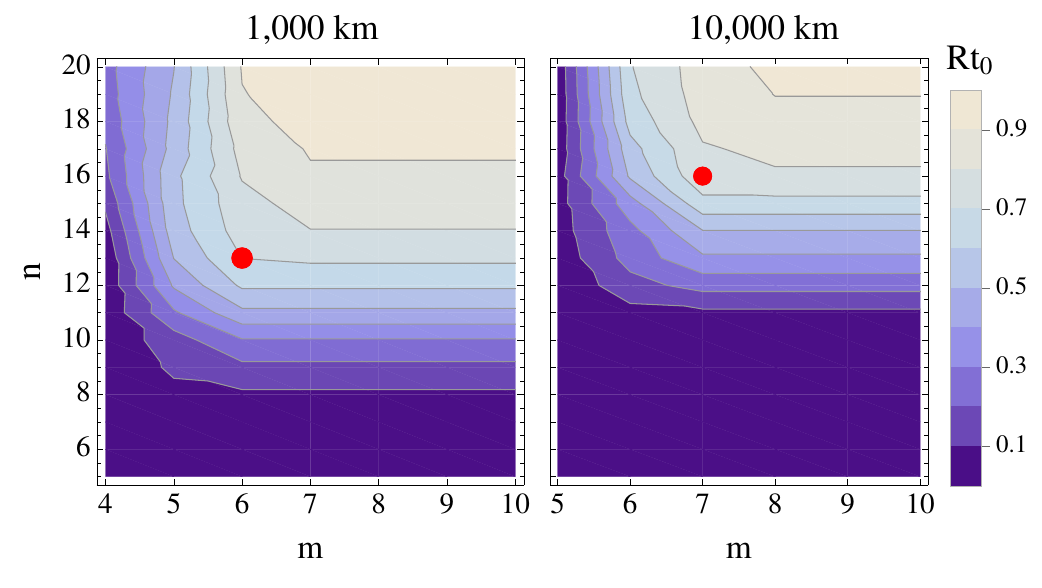}
\caption{Maximum transition probability $Rt_0$ over 1,000 and 10,000 km with repeater spacing $L_0=$1.7 km and 1.2 km, respectively, and 1\% inefficiency in each repeater ($\eta_0=0.99$). The red circle indicates the optimal choice for minimum cost $Q_{\rm min}$: $(n,m)=(13,6)$ for 1,000, and $(16,7)$ for $10,000$ km.
}\label{fig:LDSuccP}
\end{figure}

The total success probability of transmission can be obtained by $P_s(\eta_{L_0},\eta_0)^{L/L_0}\equiv Rt_0$, where $R$ is the transmission rate and $t_0$ is the time taken in the repeater. The maximum transmission probabilities over 1,000 and 10,000 km are plotted in Fig.~\ref{fig:LDSuccP}. It shows that arbitrarily high success probability approaching to unit ($\approx1$) can be attained by increasing the encoding size $N=nm$. We optimize our protocol for the total cost of photons $Q=2nmL/Rt_0L_0$
to be minimized (see details in Appendix~\ref{asec:Nume}). The optimized results by numerical searches over $\{n,m,j,L_0\}$ with exemplary parameters are presented in Table~\ref{table:resulttable}. For example, for the transmission over 1,000 km (when $\eta_0=0.93$), the best choice of encoding parameters and the repeater spacing are $(n,m,j)=(58,8,1)$ and $L_0=1.8$ km, by which $Rt_0\sim0.7$ can be achieved with total $Q_{\rm min}=7.38\times10^5$ photons. The overall transmission fidelity is estimated as $F=0.96$ by assuming depolarizing errors as detailed in Appendix~\ref{asec:elogical}. Note that the optimized results is the same for the entanglement distribution scenario in which $L_0$ is divided into half $L_0/2$ as both qubits travel to meet in the middle. 

The transmission rate $R$ is determined by the processing time $t_0$ in the repeater. We first assume that the slowest component in the repeater is the measurement process, and it takes $t_0=10~\mu$s ($1~\mu$s) (for fair comparison with other proposals \cite{Azuma15,Muralidharan14,Munro12}). Our protocol then achieves $R\sim 70$ KHz (0.7 MHz) for 1,000 km transmission (almost the same for 10,000 km). It shows the exponential superiority over the conventional all-optical approaches such as single photon transmission ($\sim10^{-10}$ Hz for 1,000 km even with a high-repetition $10$ GHz photon source) or quantum relays \cite{Jacobs2002,Riedmatten2004}. Compared to the standard repeater protocols \cite{Sangouard11,Briegle98,DLCZ,Dur99,Kok2003,Simon2007}, it is several (at least 5 to 6) order of magnitude faster. Remarkably, it also outperforms recent advanced matter-based \cite{Munro12} and all-optical based schemes~\cite{Azuma15}; it costs only an order of magnitude less ($\sim18\%$) photons to reach comparable speeds with their maximum performance \cite{Munro12,Azuma15}, and achieves nearly unit transmission probability ($Rt_0\sim1$) if the same amount of photons are used (Details of the comparison with other proposals are in Appendix~\ref{asec:compR}). 

In addition, ultrafast communications with rates up to or beyond GHz may also be expected. Note that the operations required in CBM are linear optics and photon detections, and the necessary feedforward is two or three steps (as $j=1$ or 2 in the Table~\ref{table:resulttable}) of wave plate operations. It may be thus suitable to be structured by integrated optics, possibly enabling (sub)nanosecond operation time \cite{Prevedel2007,Silverstone2014,Najafi2015,Carolan2015,Silverston16,Rudolph2016}. In our analysis, we also considered the effects of losses and imperfections on both qubits during the time for producing $\ket{\Phi^+}$ from photon pairs, but more efficient and faster generation may be expected with on-demand entangled photon sources \cite{Schwartz2016,Collins2013,Silverstone2015,Pichler2017,Buterakos2017}.

\begin{table}
\caption{\label{table:resulttable} Optimal strategies for the transmission over 1,000, 5,000, 10,000 km to minimize the total cost of photons $Q_{\rm min}$. Here, $\eta_0$ denotes the loss rate in the repeater, $Rt_0$ and $F$ are the overall transmission probability and fidelity, $(n,m,j)$ and $L_0$ are the optimal encoding and repeater spacing, respectively.}
\begin{ruledtabular}
\begin{tabular}{lccccrc}
$L$(km)&$\eta_0$
&$Q_{\rm min}$&$Rt_0$&$F$&$(n, m, j)$&$L_0$(km)\\
\hline
1,000 &0.99 & $1.3\times10^5$ &0.70 &0.98 &(13, 6, 2)&1.7 \\
&0.93 & $7.4\times10^5$ &0.70 &0.96 &(58, 8, 1)  &1.8 \\
\hline
5,000  &0.99 & $1.0\times10^6$ &0.78 &0.97 &(16, 7, 2) &1.4 \\
&0.93 & $7.4\times10^6$ &0.67 &0.93 &(83, 9, 1) &1.5 \\
\hline
10,000 &0.99 & $2.4\times10^6$ &0.77 &0.97 &(16, 7, 2) &1.2 \\
 &0.93 & $1.9\times10^7$ &0.70 &0.92 &(92, 10, 2) &1.4 \\
\end{tabular}
\end{ruledtabular}
\end{table}


\section{Conclusion} 
\label{sec:con}

We have derived the fundamental limits of the Bell measurement with linear optics and arbitrary $N$-photon encoding. First, we have proved that the success probability of the Bell measurement has the upper bound $1-2^{-N}$ by linear optics, which is the generalization of the 50\% limit of the Bell measurement with single photons ($N=1$) \cite{Calsa2001}. We have also shown that the loss-tolerance of Bell measurement (with any error correction scheme) is fundamentally limited by $\eta\eta'>0.5$ due to the no-cloning theorem, when two input qubits experience losses with rate $\eta$ and $\eta'$. These two limits of the Bell measurement determine the ultimate limit of all-optical scalability in quantum communication. 

Multiple-photon encoding in a single mode and/or Bell measurement with number resolving detection might be considered for further extension. However, we note that additional resources in Bell measurement enable us to exceed the 50\% limit \cite{Jeong2001,SLee13,Munro05,Barrett05,Ralph15}. Moreover, a resource enabling photon number resolving detection (e.g.~nonlinearity) \cite{Munro05} would, in principle, allow us to fully discriminate the Bell states \cite{Barrett05,Ralph15}. In this sense, no fundamental limit exists on the Bell measurement with arbitrary general detectors with unlimited resources. 
In order to properly generalize the 50\% limit by linear optics, the detector should thus meet the requirement of the standard Bell measurement setup (see Appendix~\ref{asec:dreso}) \cite{Weinfurter94,Calsa2001}. Therefore, $1-2^{-N}$ is the proper upper bound that is generally valid for any linear optical Bell measurements with arbitrary $N$ number of photons (note that 50\% limit is recovered when $N=1$).

We have then proposed a Bell measurement scheme with linear optics in a concatenated manner, referred to as CBM, which enables us to reach both fundamental limits by optimization. Remarkably, it outperforms all the existing Bell measurement schemes \cite{Grice2011,SLee13,Zaidi2013,Ewert2014,SLee15,SLee15a} regarding the efficiency and the loss-tolerance. 
Finally, we have constructed a building block for quantum networks based on CBM, overcoming the major obstacles of all-optical scalability. Notably, the quantum repeater based on CBM does not require long-lived quantum memories, photon-matter interactions, nor complicated circuit operations. Our protocol exhibits superiority over not only the standard quantum repeater model \cite{Sangouard11,Briegle98,DLCZ,Dur99,Kok2003,Simon2007} but also the recent advanced repeater protocols without necessitating long-lived quantum memories \cite{Munro12,Azuma15} with respect to the communication rate and resource cost.

Our work addresses both the limits and potentials of all-optical scalability as an alternative route towards long distance quantum communication. While the conventional route based on the standard quantum repeater model \cite{Sangouard11,Briegle98,DLCZ,Dur99,Kok2003,Simon2007} relies more on the development of the platforms for light-matter interaction and long-lived quantum memories \cite{Yang16,Distante17,Pu17}, our protocol, by removing the necessity of all the other demanding technologies, puts more weight on the photon sources. The major challenge for implementing our protocol is thus the preparation of large, multi-photon entangled encoded states. The recent progress of the technologies of on-demand photon sources \cite{Schwartz2016,Collins2013,Silverstone2015,Pichler2017,Buterakos2017} and platforms with integrated optics \cite{Prevedel2007,Silverstone2014,Najafi2015,Carolan2015,Silverston16,Rudolph2016} may enhance the feasibility of our protocol; conversely, our protocol, outperforming all the existing protocols (in both routes), may provide further motivations for the ongoing developments of these technologies. We expect experimental demonstrations of CBM in the near future since CBM requires only passive linear optics and photon detectors once entangled photons are prepared \cite{Pan2012,Wang2016}. Proof-of-principle tests of our repeater model are also expected in small scale network along with the progress of the abovementioned technologies. 

We emphasize that our result is not limited to all-optical quantum communication but generally applicable and valid for any quantum information protocols using Bell measurement on photons \cite{Pirandola2015,QiChao2016,Valivarthi2016}. Our scheme may be useful for designing complicated quantum networks having many participants or different applications, because it yields the same performance in two different designs for one-way communication and entanglement distribution. Moreover, it can be interchangeable or hybridizable with matter-based repeaters using the same flying qubits \cite{Muralidharan14,Munro12} in a single network design. Further studies on repeater architectures \cite{Guha2015,Pant2017,Pant2019} and other applications such as fault-tolerant quantum computation are expected.

\acknowledgments
S.W.L. thanks Koji Azuma for insightful discussions. This work was supported by the National Research Foundation of Korea (NRF) through grants funded by the Korea government (MSIP) (Grant Nos. NRF-2019R1H1A3079890 and NRF-2019M3E4A1080074). T.C.R. was supported by the Australian Research Council Centre of Excellence for Quantum Computation and Communication Technology (Project number CE170100012).

\appendix
\section{Detection resolution for the proof of the linear optical limit}
\label{asec:dreso}

\begin{figure}
\centering
\includegraphics[width=3.4in]{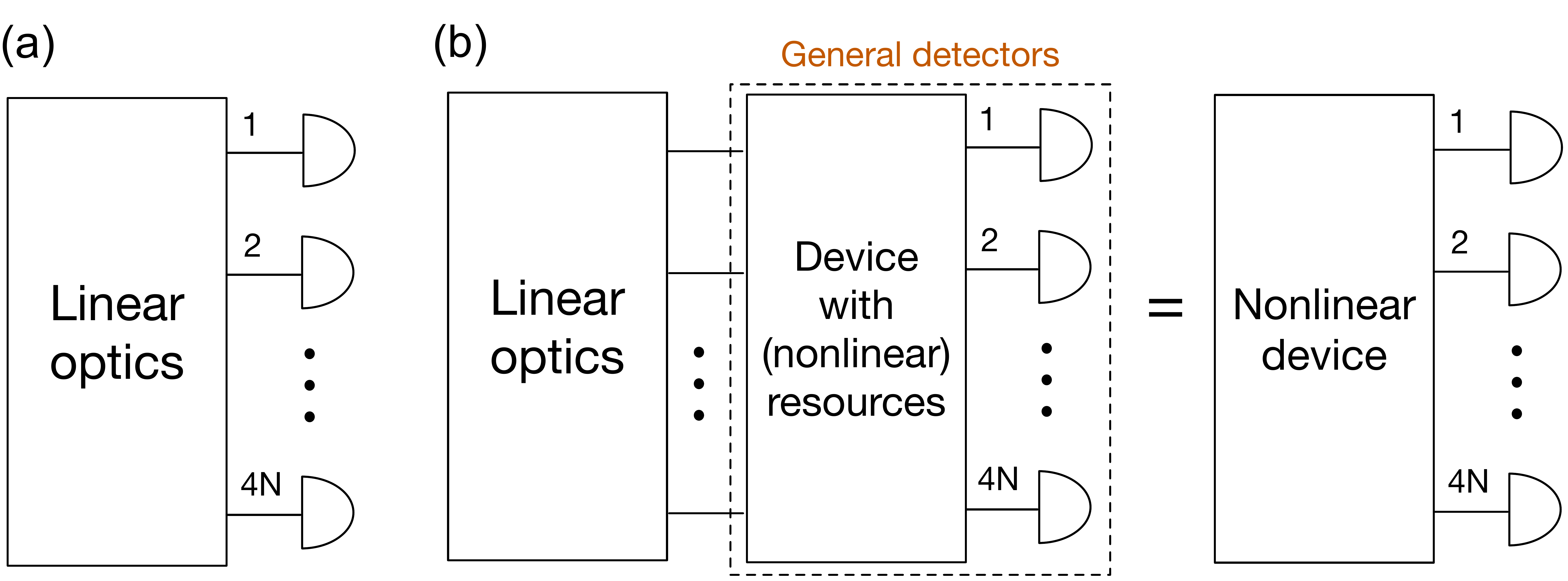}
\caption{(a) Bell measurement setup with a linear optical device and detectors resolving up to 2 photons. (b) Bell measurement setup with a linear optical device and general detectors (e.g.~photon number resolving detector) with additional resources (e.g.~nonlinearity). This is generally equivalent with a setup using arbitrary nonlinear device.
}\label{fig:SetupComp}
\end{figure}

In our derivation of the maximal success probability of Bell measurement with linear optics, we assume that the detectors can resolve  0, 1, and $\geq 2$, to meet the minimum requirement of the standard Bell measurement technique with linear optics \cite{Weinfurter94} as illustrated as in Fig.~\ref{fig:SetupComp}(a). One might consider arbitrary general detectors such as photon number resolving detector for further extension of the limit. However, from the fact that such a detector generally requires additional resources, the detection part can be decomposed again into a device with additional resources and detectors resolving  0, 1, and $\geq 2$. This is, in turn, equivalent with the general Bell measurement setup of a device consuming additional resources plus detectors as illustrated in Fig.~\ref{fig:SetupComp}(b). In fact, a resource enabling photon number resolving detection such as nonlinearity \cite{Munro05} can, in principle, also allow us to fully discriminate Bell states \cite{Barrett05,Ralph15}. As a result, no fundamental limit would exist on the Bell measurement with arbitrary general detectors with unlimited resources. Therefore, in order to properly and fairly generalize the 50\% limit, the detection resolution should meet the requirement of the standard Bell measurement setup, i.e., (0, 1, $\geq 2$) discrimination.

\section{Limit of the success probability of BM with $N$ photons and linear optics}
\label{asec:limit}

Let us prove the maximum success probability of linear optical Bell measurements performed on logical qubits each encoded with $N$ photons. Equiprobable four logical Bell states, including total $2N$ photons in the polarization degree of freedom, are prepared to enter the input modes of the linear optical devices. They are detected at the output modes as illustrated in Fig.~\ref{fig:FBounds}(a). We can write the creation operator of the output modes $\{\hat{c}^{\dag}_j\}$ as the linear combination of the creation operators of the input modes $\{\hat{a}^{\dag}_i\}$, {\em i.e.} $\hat{c}^{\dag}_j=\sum^{4N}_iU_{ji}a^{\dag}_i$ where $U$ is the unitary matrix for the linear optical devices. The input and output mode vectors can be defined respectively as $\vec{a}=(a^{\dag}_1,\cdots,a^{\dag}_{4N})^T$ and $\vec{c}=(c^{\dag}_1,\cdots,c^{\dag}_{4N})^T$, and $\vec{\alpha}_i=(U_{i1},\cdots,U_{i4N})^T$ as the $i$th column vector of $U$. In the dual-rail representation, each photon occupies 2 modes. Like the standard Bell measurement scheme with single photons \cite{Weinfurter94,Braunstein94,Michler95}, we assume here that the detectors can resolve up to 2 photons. This is the minimum requirement to distinguish all possible outcomes in the standard Bell measurement scheme \cite{Weinfurter94,Braunstein94,Michler95} and detect losses.

For single photon encoding ($N=1$), we can follow the proof in Ref.~\cite{Calsa2001}. The input state can be written in general by $\ket{\Psi}=\sum^{4}_{i,j}N_{ij}\hat{a}^{\dag}_i\hat{a}^{\dag}_j\ket{0}=\vec{a}^T\cdot {\bf N} \cdot\vec{a}\ket{0}$, where $\bf N$ is a $4\times4$ symmetric matrix. This can be rewritten in terms of the output modes by $\ket{\Psi}=\vec{a}^T\cdot {\bf N} \cdot\vec{a}\ket{0}=\vec{c}^T\cdot {\bf M} \cdot\vec{c}\ket{0}=\sum^{4}_{k,l}M_{kl}\hat{c}^{\dag}_k\hat{c}^{\dag}_l\ket{0}$, where ${\bf M}=U^T{\bf N}U$. For the Bell states, $\ket{\Psi^{\mu=1,2}}=a^{\dag}_1a^{\dag}_3\pm a^{\dag}_2a^{\dag}_4\ket{0}$ and $\ket{\Psi^{\mu=3,4}}=a^{\dag}_1a^{\dag}_4\pm a^{\dag}_2a^{\dag}_3\ket{0}$ (coefficient omitted hereafter), ${\bf M}^{\mu}=U^T{\bf N}^{\mu}U$ where 
\[
{\bf N}^{\mu}=\frac{1}{2\sqrt{2}}
\begin{bmatrix}
    0  & 0 & \delta_{\mu1}+\delta_{\mu2} &\delta_{\mu3}+\delta_{\mu4} \\
    0  & 0 & \delta_{\mu3}-\delta_{\mu4} &\delta_{\mu1}-\delta_{\mu2} \\
    \delta_{\mu1}+\delta_{\mu2} & \delta_{\mu3}-\delta_{\mu4}& 0  & 0 \\
    \delta_{\mu3}+\delta_{\mu4} & \delta_{\mu1}-\delta_{\mu2} & 0  & 0
\end{bmatrix}
.
\]
The contribution of the different Bell states to particular detection events can be investigated by the form of ${\bf M}^{\mu}$:

If we consider two-photon detection events, the probability that two photons are detected at mode $c_k$ for the input state $\ket{\Psi^{\mu}}$ is $P^{\mu}_k[2]=\bra{0}c^2_k M^{\mu *}_{kk} M^{\mu}_{kk} c^{\dag 2}_k \ket{0}=2|\vec{\alpha}^T_k\cdot{\bf N}^{\mu}\cdot\vec{\alpha}_k|$. In order to identify a Bell state (e.g. when $\mu=1$) by this event, the probabilities for all three other Bell states $\mu=2,3,4$ should be vanished. However, this is impossible because $P^{\mu}_k[2]=0$ for all $\mu$ with any $\vec{\alpha}_k$.

For single-photon detection at mode $c_k$, we can write the conditional state as $\ket{\Phi^{\mu}_k}=2\sum^4_{l\neq k}M^{\mu}_{kl}c^{\dag}_l\ket{0}=2(\vec{m}_k^{\mu T}\cdot\vec{c}-M^{\mu}_{kk}c^{\dag}_k)$, where $\vec{m}_k^{\mu T}=U^T{\bf N}^{\mu}\vec{\alpha}_k=U^T\vec{v}_k^{\mu}$ is the $s$th column vector of $\bf M$ with $\vec{v}^{\mu}_k={\bf N}^{\mu}\vec{\alpha}_k$. Here, the vectors $\{\vec{v}^{1}_k,\vec{v}^{2}_k,\vec{v}^{3}_k,\vec{v}^{4}_k\}$ correspond to four input Bell states so that they are linearly dependent and have the same norm $|\vec{v}^{\mu}_k|^2=|\vec{\alpha}_k|^2$, {\em i.e.} $\sum^4_{\mu=1}b_{\mu}\ket{\Phi^{\mu}_k}=0$ with at least two $b_{\mu}\neq0$. This implies that the maximum number of linearly independent ({\em i.e.}, unambiguously discriminated) $\ket{\Phi^{\mu}_k}$ is two. With the probability of single photon detection at mode $c_k$, $P^{\mu}_k[1]=\braket{\Phi^{\mu}_k}{\Phi^{\mu}_k}=(|\vec{\alpha}_k|^2-|\vec{\alpha}_k\cdot\vec{v}^{\mu}_k|^2)$, the upper bound of success probability that the detection in mode $c_k$ contributes to the unambiguous discrimination of a Bell state is obtained by $p_k \leq \frac{1}{4}(P^{\mu=a}_k[1]+P^{\mu=b}_k[1])=\frac{1}{4}|\vec{\alpha}_k|^2$. Therefore, the upper bound of the total success probability of the Bell measurement with $N=1$ can be obtained by $P_s \leq \frac{1}{2}\sum^4_{k=1}p_i=\frac{1}{8}\sum^4_{k=1}|\vec{\alpha}_k|^2=\frac{1}{8}\sum^4_{k=1}\sum^4_{l=1}|U_{kl}|^2=\frac{1}{2}$. The upper bound of the success probability is the same for the case when including ancillary modes in vacuum states \cite{Calsa2001}.

We now consider the encoding with arbitrary $N>1$ photons. The logical basis $\ket{0_L}$ and $\ket{1_L}$ are assumed to be defined without redundancy. For example, $\ket{0_L}=\ket{H}\ket{V}$ and $\ket{1_L}=\ket{V}\ket{V}$, or $\ket{0_L}=\ket{H}\ket{+}$ and $\ket{1_L}=\ket{V}\ket{V}$ are redundantly encoded with the second polarization mode. Therefore, the logical basis can be generally represented by $\ket{0_L}={\cal U}a^{\dag}_{i_1}\cdots a^{\dag}_{i_N}\ket{0}$ and $\ket{1_L}={\cal U}a^{\dag}_{j_1}\cdots a^{\dag}_{j_N}\ket{0}$, where ${\cal U}$ is an arbitrary unitary operation and $i_1,\cdots,i_N,j_1,\cdots,j_N$ are the mode numbers given by a permutation of 1 to 2N.

For two-photon encoding ($N=2$), the logical basis are generally $\ket{0_L}={\cal U}a^{\dag}_{i_1}a^{\dag}_{i_2}\ket{0}$ and $\ket{1_L}={\cal U}a^{\dag}_{j_1}a^{\dag}_{j_2}\ket{0}$, with a permutation $(i_1,i_2,j_1,j_2)$ of 1 to 4. For example, one may choose $\ket{0_L}=\cos{\theta}\ket{H}\ket{H}-\sin{\theta}\ket{V}\ket{V}$ and $\ket{1_L}=\sin{\theta}\ket{H}\ket{H}+\cos{\theta}\ket{V}\ket{V}$ with $\theta\in[0,\pi/2]$, or their variations with local unitary operations. The logical Bell states can be then written by $\ket{\Psi^{\mu=1,2}}={\cal U}{\cal U'}(a^{\dag}_{i_1}a^{\dag}_{i_2}a^{\dag}_{i'_1}a^{\dag}_{i'_2}\pm a^{\dag}_{j_1}a^{\dag}_{j_2}a^{\dag}_{j'_1}a^{\dag}_{j'_2})\ket{0}$ and $\ket{\Psi^{\mu=3,4}}={\cal U}{\cal U'}(a^{\dag}_{i_1}a^{\dag}_{i_2}a^{\dag}_{j'_1}a^{\dag}_{j'_2}\pm a^{\dag}_{j_1}a^{\dag}_{j_2}a^{\dag}_{i'_1}a^{\dag}_{i'_2})\ket{0}$. Without loss of the generality, these can be rearranged as $\ket{\Psi^{\mu=1,2}}={\cal U}{\cal U'}(a^{\dag}_{i_1}a^{\dag}_{i_1'}a^{\dag}_{i_2}a^{\dag}_{i'_2}\pm a^{\dag}_{j_1}a^{\dag}_{j'_1}a^{\dag}_{j_2}a^{\dag}_{j_2'})\ket{0}$ and $\ket{\Psi^{\mu=3,4}}={\cal U}{\cal U'}(a^{\dag}_{i_1}a^{\dag}_{j_1'}a^{\dag}_{i_2}a^{\dag}_{j_2'}\pm a^{\dag}_{j_1}a^{\dag}_{i_1'}a^{\dag}_{j_2}a^{\dag}_{i_2'})\ket{0}$, and then rewritten by 
\begin{equation}
\begin{aligned}
\ket{\Psi^{1}}&={\cal U}{\cal U'}(\big[(a^{\dag}_{i_1}a^{\dag}_{i_1'}+a^{\dag}_{j_1}a^{\dag}_{j_1'})(a^{\dag}_{i_2}a^{\dag}_{i_2'}+a^{\dag}_{j_2}a^{\dag}_{j_2'})\\
\nonumber
&\hspace{10mm}+(a^{\dag}_{i_1}a^{\dag}_{i_1'}-a^{\dag}_{j_1}a^{\dag}_{j_1'})(a^{\dag}_{i_2}a^{\dag}_{i_2'}-a^{\dag}_{j_2}a^{\dag}_{j_2'})\big]\ket{0}\\
\ket{\Psi^{2}}&={\cal U}{\cal U'}(\big[(a^{\dag}_{i_1}a^{\dag}_{i_1'}+a^{\dag}_{j_1}a^{\dag}_{j_1'})(a^{\dag}_{i_2}a^{\dag}_{i_2'}-a^{\dag}_{j_2}a^{\dag}_{j_2'})\\
\nonumber
&\hspace{10mm}+(a^{\dag}_{i_1}a^{\dag}_{i_1'}-a^{\dag}_{j_1}a^{\dag}_{j_1'})(a^{\dag}_{i_2}a^{\dag}_{i_2'}+a^{\dag}_{j_2}a^{\dag}_{j_2'})\big]\ket{0}\\
\ket{\Psi^{3}}&={\cal U}{\cal U'}(\big[(a^{\dag}_{i_1}a^{\dag}_{j_1'}+a^{\dag}_{j_1}a^{\dag}_{i_1'})(a^{\dag}_{i_2}a^{\dag}_{j_2'}+a^{\dag}_{j_2}a^{\dag}_{i_2'})\\
\nonumber
&\hspace{10mm}+(a^{\dag}_{i_1}a^{\dag}_{j_1'}-a^{\dag}_{j_1}a^{\dag}_{i_1'})(a^{\dag}_{i_2}a^{\dag}_{j_2'}-a^{\dag}_{j_2}a^{\dag}_{i_2'})\big]\ket{0}\\
\ket{\Psi^{4}}&={\cal U}{\cal U'}(\big[(a^{\dag}_{i_1}a^{\dag}_{j_1'}+a^{\dag}_{j_1}a^{\dag}_{i_1'})(a^{\dag}_{i_2}a^{\dag}_{j_2'}-a^{\dag}_{j_2}a^{\dag}_{i_2'})\\
\nonumber
&\hspace{10mm}+(a^{\dag}_{i_1}a^{\dag}_{j_1'}-a^{\dag}_{j_1}a^{\dag}_{i_1'})(a^{\dag}_{i_2}a^{\dag}_{j_2'}+a^{\dag}_{j_2}a^{\dag}_{i_2'})\big]\ket{0}.
\end{aligned}
\end{equation}
Therefore, all the input states are generally represented as $\ket{\Psi^{\mu}}=\sum_{\nu}{\cal C}^{\mu}_{\nu}\ket{\psi^{\nu}}$, where $\sum_{\nu}|{\cal C}_{\nu}|^2=1$, i.e., a linear combinations of 
\begin{equation}
\label{eq:Bellform}
\begin{aligned}
\ket{\psi^{\nu}}&=\sum^2_{p=1}\Big(\sum^{4}_{i,j=1}N^{\nu,p}_{ij}\hat{a}^{\dag}_i\hat{a}^{\dag}_j\Big)_1\Big(\sum^8_{i',j'=5}N'^{\nu,p}_{i'j'}\hat{a}^{\dag}_{i'}\hat{a}^{\dag}_{j'}\Big)_2\ket{0}\\&=\sum^2_{p=1}\big(\vec{a}^T\cdot {\bf N}^{\nu,p} \cdot\vec{a}\big)_1\big(\vec{a'}^T\cdot {\bf N'}^{\nu,p} \cdot\vec{a'}\big)_2\ket{0}.
\end{aligned}
\end{equation}

Let us first consider the case when the Bell states are encoded exactly in the form of $\ket{\Psi^{\mu}}=\ket{\psi^{\mu}}$ given in (\ref{eq:Bellform}) with ${\cal C}_{\nu=\mu}=1$, i.e.,~${\cal U}={\cal U'}=I$. If we consider the unitary linear-optical map on the first block of $\ket{\psi^{\mu}}$, the output modes $\hat{c}^{\dag}_k=\sum^{4}_{i=1}U_{ki}a^{\dag}_i$ can be arranged to be labeled as $k=1,2,3,4$. In this configuration only, the possible detection event at the output modes $c_k$ are either two-photon detection or single photon detection by which the input state of the first block can be read out. However, if the unitary linear-optical map on the second block $\hat{c}^{\dag}_l=\sum^{8}_{i=5}U_{li}a^{\dag}_i$ shares any output mode with the map on the first block (i.e. $c_k=c_l$), the discrimination of the input states becomes obviously harder as the detectors can resolve up to 2 photons. This is equivalent effectively with the loss of the surplus photons. We can thus restrict the unitary matrix for the linear optical device to be decomposed into two $4\times4$ unitary matrices applying separately to the first and second block, $U=U_1\otimes U_2$, in order to evaluate the maximum success probability.

For the events that a photon from the first block is detected at mode $c_k$ and a photon from the second block is detected at mode $c_l$, we can write the conditional state as $\ket{\Phi^{\mu}_{kl}}=\bra{0}c_k c_l \ket{\psi^{\mu}}=\sum^2_{p=1}\ket{\phi^{\mu,p}_k}\ket{\phi^{\mu,p}_l}$, where $\ket{\phi^{\mu,p}_k}=2\sum^4_{i=1}M^{\mu,p}_{ki}c^{\dag}_i\ket{0}=2\vec{m}_k^{\mu,p T}\cdot\vec{c}_k\ket{0}$ with $\vec{m}_k^{\mu,p T}=U_1^T{\bf N}^{\mu,p}\vec{\alpha}_k=U_1^T\vec{v}_k^{\mu,p}$, and $\ket{\phi^{\mu,p}_l}=2\sum^8_{j=5}M'^{\mu,p}_{lj}c^{\dag}_j\ket{0}=2\vec{m}_l^{\mu,p T}\cdot\vec{c}_l\ket{0}$ with $\vec{m}_l^{\mu,p T}=U_2^T{\bf N'}^{\mu,p}\vec{\alpha}_l=U_1^T\vec{v}_l^{\mu,p}$. The vectors $\{\vec{v}^{1,p}_k,\vec{v}^{2,p}_k,\vec{v}^{3,p}_k,\vec{v}^{4,p}_k\}$ have the same norm $|\vec{v}^{\mu,p}_k|^2=|\vec{\alpha}_k|^2$ and are linearly dependent so that $\sum^4_{\mu=1}b^{\mu,p}_k\ket{\phi^{\mu,p}_k}=0$ with at least two $b^{\mu,p}_k\neq0$. Likewise, for the vectors $\{\vec{v}^{1,p}_l,\vec{v}^{2,p}_l,\vec{v}^{3,p}_l,\vec{v}^{4,p}_l\}$, $\sum^4_{\mu=1}b^{\mu,p}_l\ket{\phi^{\mu,p}_l}=0$ with at least two $b^{\mu,p}_l\neq0$. Thus, the conditional states $\ket{\Phi^{\mu}_{kl}}$ for different $\mu$ are linearly dependent $\sum^4_{\mu=1}b_{\mu}\ket{\Phi^{\mu}_{kl}}=0$ with $b_{\mu}\neq0$ only when both $b^{\mu,p}_k\neq0$ and $b^{\mu,p}_l\neq0$ for $p=1,2$. In other words, $\ket{\Phi^{\mu=a}_{kl}}$ is linearly independent to others as long as either $b^{\mu,p}_k=0$ or $b^{\mu,p}_l=0$ for any $p$ so that the input Bell state can be unambiguously discriminated. The maximum probability that the detection at mode $c_k$($c_l$) contributes to the unambiguous discrimination in first(second) block is $p_k=\frac{1}{4}|\vec{\alpha}_k|^2$($p_l=\frac{1}{4}|\vec{\alpha}_l|^2$). The maximum total success probability to distinguish the input states can be then obtained by $P_s \leq 1-(1-\frac{1}{2}\sum^4_{k=1}p_k)(1-\frac{1}{2}\sum^8_{l=5}p_l)=\frac{1}{2}(\sum^4_{k=1}p_k+\sum^8_{l=5}p_l)-\frac{1}{4}\sum^4_{k=1}p_k\sum^8_{l=5}p_l=1-\frac{1}{4}=\frac{3}{4}$.

This upper bound is generally valid for arbitrary input Bell states $\ket{\Psi^{\mu}}=\sum_{\nu}{\cal C}^{\mu}_{\nu}\ket{\psi^{\nu}}$, i.e.~arbitrary ${\cal U}$ and ${\cal U'}$. In this case, the conditional state, for the events that a photon from the first(second) block is detected at mode $c_k$($c_l$), is written by $\ket{\Phi^{\mu}_{kl}}=\bra{0}c_k c_l \ket{\Psi^{\mu}}=\sum_\nu{\cal C}^{\mu}_{\nu}\sum^2_{p=1}\ket{\phi^{\nu,p}_k}\ket{\phi^{\nu,p}_l}$. Since $\ket{\Phi^{\mu}_{kl}}$ is a linear combination of $\sum^2_{p=1}\ket{\phi^{\nu,p}_k}\ket{\phi^{\nu,p}_l}$, the number of linearly independent $\ket{\Phi^{\mu}_{kl}}$ is at best the same with the number of linearly independent $\sum^2_{p=1}\ket{\phi^{\nu,p}_k}\ket{\phi^{\nu,p}_l}$. Therefore, the success probability to unambiguously discriminate $\ket{\Psi^{\mu=1,2,3,4}}$ is upper bounded by the success probability to unambiguously discriminate $\ket{\psi^{\mu=1,2,3,4}}$, i.e., $P_s(\ket{\Psi^{\mu}})\leq P_s(\ket{\psi^{\mu}})\leq\frac{3}{4}$.

It is straightforward to extend the proof to arbitrary $N$, as the input Bell states can be written by $\ket{\Psi^{\mu}}=\sum_{\nu}{\cal C}^{\mu}_{\nu}\sum^2_{p=1}\bigotimes^N_{q=1} \Big(\sum^{4q}_{i_q,j_q=4q-3}N^{\nu,p}_{i_q j_q}\hat{a}^{\dag}_{i_q}\hat{a}^{\dag}_{j_q}\Big)\ket{0}$, and the matrix of linear-optical unitary map can be restricted to the form of $U=\bigotimes_{q=1}^N U_q$ to evaluate the maximum success probability. Finally, the upper bound of the success probability is obtained by $P_s(\ket{\Psi^{\mu}}) \leq 1-\prod^{N}_{q=1}(1-\frac{1}{2}\sum^{4q}_{i_q=4q-3}p_{i_q})=1-\prod^N_{q=1}\frac{1}{2}=1-2^{-N}$.

\section{Decomposition of the encoded Bell states}  
\label{asec:Decom}

In the parity state encoding, the logical basis are defined as $\ket{0_L}=\ket{+^{(m)}}_1\cdots\ket{+^{(m)}}_n$ and $\ket{1_L}=\ket{-^{(m)}}_1\cdots\ket{-^{(m)}}_n$ with $\ket{\pm^{(m)}}=(\ket{H}_1\cdots\ket{H}_m\pm\ket{V}_1\cdots\ket{V}_m)/\sqrt{2}$, and the Bell states are written by
\begin{equation}
\nonumber
\begin{aligned}
\label{eq:logicalBellSA}
\ket{\Phi^{\pm}}&=\frac{1}{\sqrt{2}}\Big(\ket{+^{(m)}}_1\cdots\ket{+^{(m)}}_n \ket{+^{(m)}}_{1'}\cdots\ket{+^{(m)}}_{n'}\\
&\hspace{10mm}\pm\ket{-^{(m)}}_1\cdots\ket{-^{(m)}}_n \ket{-^{(m)}}_{1'}\cdots\ket{-^{(m)}}_{n'}\Big)\\
\ket{\Psi^{\pm}}&=\frac{1}{\sqrt{2}}\Big(\ket{+^{(m)}}_1\cdots\ket{+^{(m)}}_n \ket{-^{(m)}}_{1'}\cdots\ket{-^{(m)}}_{n'}\\
&\hspace{10mm}\pm\ket{-^{(m)}}_1\cdots\ket{-^{(m)}}_n \ket{+^{(m)}}_{1'}\cdots\ket{+^{(m)}}_{n'}\Big),
\end{aligned}
\end{equation}
where the first $n$ blocks (from $1$ to $n$) are from the first qubit and the following $n$ blocks (from $1'$ to $n'$) are from the second qubit. By rearranging the order of blocks $(1,\ldots,n,1',\ldots,n')$ to $(1,1',2,2',\ldots,n,n')$, these can be completely decomposed into the 1st (block) level Bell states. For example, 
\begin{equation}
\nonumber
\begin{aligned}
\label{eq:level2Bell1a}
\ket{\Phi^{+}} &= \frac{1}{\sqrt{2}}\Big(\ket{+^{(m)}}_1\cdots\ket{+^{(m)}}_n \ket{+^{(m)}}_{1'}\cdots\ket{+^{(m)}}_{n'}\\
&\hspace{10mm}+\ket{-^{(m)}}_1\cdots\ket{-^{(m)}}_n \ket{-^{(m)}}_{1'}\cdots\ket{-^{(m)}}_{n'}\Big)\\
&= \frac{1}{\sqrt{2}}\Big(\ket{+^{(m)}}_1\ket{+^{(m)}}_{1'}\cdots\ket{+^{(m)}}_n\ket{+^{(m)}}_{n'}\\
&\hspace{10mm}+\ket{-^{(m)}}_1\ket{-^{(m)}}_{1'}\cdots\ket{-^{(m)}}_n\ket{-^{(m)}}_{n'}\Big)\\
&=\frac{1}{\sqrt{2^{n-1}}}\Big(\ket{\phi^+_{(m)}}_{11'}\ket{\phi^+_{(m)}}_{22'}\cdots\ket{\phi^+_{(m)}}_{nn'}\\
&+\ket{\phi^-_{(m)}}_{11'}\ket{\phi^-_{(m)}}_{22'}\ket{\phi^+_{(m)}}_{33'}\cdots\ket{\phi^+_{(m)}}_{nn'}\\
&+\ket{\phi^-_{(m)}}_{11'}\ket{\phi^+_{(m)}}_{22'}\ket{\phi^-_{(m)}}_{33'}\cdots\ket{\phi^+_{(m)}}_{nn'}\\
&\hspace{24mm}\vdots\\
&+\ket{\phi^+_{(m)}}_{11'}\cdots\ket{\phi^+_{(m)}}_{n-2,n-2'}\ket{\phi^-_{(m)}}_{n-1,n-1'}\ket{\phi^-_{(m)}}_{nn'}\\
&+\cdots\Big)\\
&=\frac{1}{\sqrt{2^{n-1}}}\sum_{j={\rm even} \leq n}{\cal P}[\ket{\phi^-_{(m)}}^{\otimes j}\ket{\phi^+_{(m)}}^{\otimes n-j}],
\end{aligned}
\end{equation}
which is the equally weighted superposition of all possible $n$-fold tensor products of even number $j$ of $\ket{\phi^-_{(m)}}$ and $n-j$ of $\ket{\phi^+_{(m)}}$. Likewise for others, all the (logical) 2nd level Bell states can be represented by
\begin{equation}
\nonumber
\begin{aligned}
\label{eq:level2Bellothersa}
&\ket{\Phi^{+(-)}}=\frac{1}{\sqrt{2^{n-1}}}\sum_{j={\rm even(odd)} \leq n}{\cal P}[\ket{\phi^-_{(m)}}^{\otimes j}\ket{\phi^+_{(m)}}^{\otimes n-j}],\\
&\ket{\Psi^{+(-)}}=\frac{1}{\sqrt{2^{n-1}}}\sum_{j={\rm even(odd)} \leq n}{\cal P}[\ket{\psi^-_{(m)}}^{\otimes j}\ket{\psi^+_{(m)}}^{\otimes n-j}],
\end{aligned}
\end{equation}
where ${\cal P}[\cdot]$ is defined as a permutation function e.g. ${\cal P}[\ket{\phi^-}\ket{\phi^+}\ket{\phi^+}]=\ket{\phi^-}\ket{\phi^+}\ket{\phi^+}+\ket{\phi^+}\ket{\phi^-}\ket{\phi^+}+\ket{\phi^+}\ket{\phi^+}\ket{\phi^-}$. Note that $\ket{\Phi^{+(-)}}$ includes even (odd) number of $\ket{\phi^-_{(m)}}$, while $\ket{\Psi^{+(-)}}$ includes even (odd) number of $\ket{\psi^-_{(m)}}$.

The 1st level Bell states, $\ket{\phi^{\pm}_{(m)}}=(\ket{+^{(m)}}\ket{+^{(m)}}\pm\ket{-^{(m)}}\ket{-^{(m)}})/\sqrt{2}$ and $\ket{\phi^{\pm}_{(m)}}=(\ket{+^{(m)}}\ket{+^{(m)}}\pm\ket{-^{(m)}}\ket{-^{(m)}})/\sqrt{2}$, can be written again as
\begin{equation}
\nonumber
\begin{aligned}
\label{eq:level1Bells-2}
&\ket{\phi^{+}_{(m)}}=(\ket{H}_1\cdots\ket{H}_m\ket{H}_{1'}\cdots\ket{H}_{m'}\\
&\hspace{20mm}+\ket{V}_1\cdots\ket{V}_m\ket{V}_{1'}\cdots\ket{V}_{m'})/\sqrt{2},\\
&\ket{\phi^{-}_{(m)}}=(\ket{H}_1\cdots\ket{H}_m\ket{V}_{1'}\cdots\ket{V}_{m'}\\
&\hspace{20mm}+\ket{V}_1\cdots\ket{V}_m\ket{H}_{1'}\cdots\ket{H}_{m'})/\sqrt{2},\\
&\ket{\psi^{+}_{(m)}}=(\ket{H}_1\cdots\ket{H}_m\ket{H}_{1'}\cdots\ket{H}_{m'}\\
&\hspace{20mm}-\ket{V}_1\cdots\ket{V}_m\ket{V}_{1'}\cdots\ket{V}_{m'})/\sqrt{2},\\
&\ket{\psi^{-}_{(m)}}=(\ket{H}_1\cdots\ket{H}_m\ket{V}_{1'}\cdots\ket{V}_{m'}\\
&\hspace{20mm}-\ket{V}_1\cdots\ket{V}_m\ket{H}_{1'}\cdots\ket{H}_{m'})/\sqrt{2},
\end{aligned}
\end{equation}
by $\ket{\pm^{(m)}}=(\ket{H}_1\cdots\ket{H}_m\pm\ket{V}_1\cdots\ket{V}_m)/\sqrt{2}$. By rearranging the order of modes $(1,\ldots,m,1',\ldots,m')$  to $(1,1',2,2',\ldots,m,m')$, these are similarly decomposed into the 0th level Bell states, $\ket{\phi^{\pm}}=(\ket{+}\ket{+}\pm\ket{-}\ket{-})/\sqrt{2}$ and $\ket{\psi^{\pm}}= (\ket{+}\ket{-}\pm\ket{-}\ket{+})/\sqrt{2}$ where $\ket{\pm}=(\ket{H}\pm\ket{V})/\sqrt{2}$. For example,  
\begin{equation}
\begin{aligned}
\nonumber
\label{eq:level1Bell1}
\ket{\phi^{+}_{(m)}}&=(\ket{H}_1\cdots\ket{H}_m\ket{H}_{1'}\cdots\ket{H}_{m'}\\
&\hspace{10mm}+\ket{V}_1\cdots\ket{V}_m\ket{V}_{1'}\cdots\ket{V}_{m'})/\sqrt{2}\\
&=(\ket{H}_1\ket{H}_{1'}\cdots\ket{H}_m\ket{H}_{m'}\\
&\hspace{10mm}+\ket{V}_1\ket{V}_{1'}\cdots\ket{V}_m\ket{V}_{m'})/\sqrt{2}\\
&=\frac{1}{\sqrt{2^{m-1}}}\Big(\ket{\phi^+}_{11'}\ket{\phi^+}_{22'}\cdots\ket{\phi^+}_{mm'}\\
&+\ket{\psi^+}_{11'}\ket{\psi^+}_{22'}\ket{\phi^+}_{33'}\cdots\ket{\phi^+}_{mm'}\\
&+\ket{\phi^+}_{11'}\ket{\psi^+}_{22'}\ket{\psi^+}_{33'}\cdots\ket{\phi^+}_{mm'}\\
&\hspace{24mm}\vdots\\
&+\ket{\phi^+}_{11'}\cdots\ket{\phi^+}_{m-2,m-2'}\ket{\psi^+}_{m-1,m-1'}\ket{\psi^+}_{mm'}\\
&+\cdots\Big)\\
&=\frac{1}{\sqrt{2^{m-1}}}\sum_{k={\rm even} \leq m}{\cal P}[\ket{\psi^+}^{\otimes k}\ket{\phi^+}^{\otimes m-k}],
\end{aligned}
\end{equation}
in the form of the equally weighted superposition of all possible $m$-fold tensor products of even number $k$ of $\ket{\psi^+}$ and $m-k$ of $\ket{\phi^+}$. Likewise for others, we can rewrite all the 1st level Bell states as
\begin{equation}
\begin{aligned}
\label{eq:level1Bellothers}
&\ket{\phi^{\pm}_{(m)}}=\frac{1}{\sqrt{2^{m-1}}}\sum_{k={\rm even} \leq m}{\cal P}[\ket{\psi^{\pm}}^{\otimes k}\ket{\phi^{\pm}}^{\otimes m-k}],\\
&\ket{\psi^{\pm}_{(m)}}=\frac{1}{\sqrt{2^{m-1}}}\sum_{k={\rm odd} \leq m}{\cal P}[\ket{\psi^{\pm}}^{\otimes k}\ket{\phi^{\pm}}^{\otimes m-k}].
\end{aligned}
\end{equation}
Note that $\ket{\phi^{\pm}_{(m)}}$ includes even number of $\ket{\psi^{\pm}}$, while $\ket{\psi^{\pm}_{(m)}}$ includes odd number of $\ket{\psi^{\pm}}$.

\section{CBM scheme}
\label{sec:detailCBM}

\begin{figure}
\centering
\includegraphics[width=3.2in]{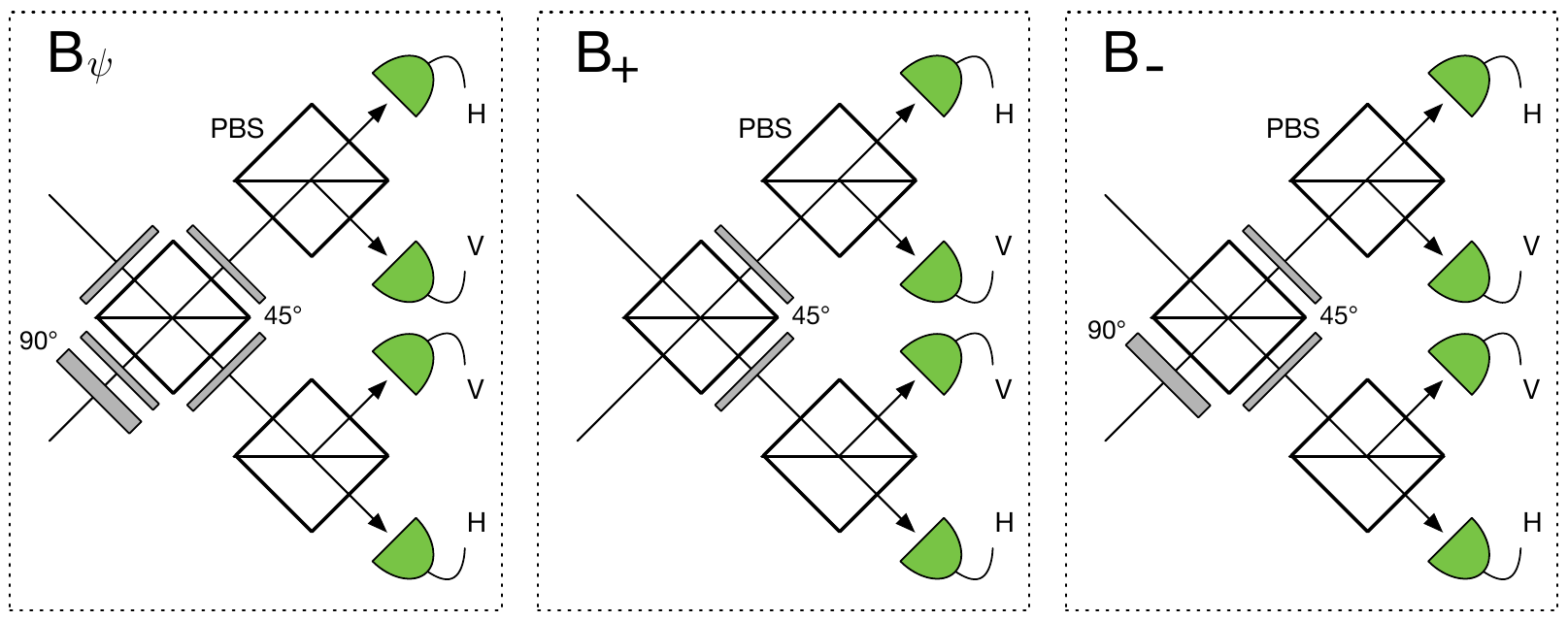}
\caption{Three types of 0th level Bell measurements on photon pairs as ${\rm B}_{(0)}=\{{\rm B}_{\psi}, {\rm B}_{+}, {\rm B}_{-}\}$, discriminating ($\ket{\psi^+}, \ket{\psi^-}$, $\ket{\phi^+},\ket{\psi^+}$, $\ket{\phi^-},\ket{\psi^-}$) respectively, and convertible by  changing the wave plates at the input modes of the first PBS.
}\label{fig:aB0scheme}
\end{figure}

{\bf (0th level)} For the Bell measurement on photon pairs (denoted as 0th level Bell measurement ${\rm B}_{(0)}$), we basically employ the standard technique using linear optical elements such as polarizing beam splitter (PBS), wave plates and photon detection , unambiguously discriminating two of the four Bell states $\ket{\phi^{\pm}}$ and $\ket{\psi^{\pm}}$. We use three different types (${\rm B}_{\psi}$, ${\rm B}_{+}$, ${\rm B}_{-}$), which respectively discriminates ($\ket{\psi^+},\ket{\psi^-}$, $\ket{\phi^+},\ket{\psi^+}$, $\ket{\phi^-},\ket{\psi^-}$) as illustrated in Fig.~\ref{fig:aB0scheme}. The two identified Bell states out of four can be chosen by changing the wave plates at the input modes of the first PBS. For example, ${\rm B}_{\psi}$ yields the following outcomes: (Success) if $\ket{\psi^+}$ and $\ket{\psi^-}$ state enter into $\rm B_{\psi}$, at the first PBS two photons are separated into different modes resulting in one click from the upper two detectors and another from lower two. From all possible events of separated clicks, $\ket{\psi^+}$ and $\ket{\psi^-}$ can be deterministically identified: (H,H)~or~(V,V) click for $\ket{\psi^+}$, and (H,V)~or~(V,H) for $\ket{\psi^-}$. (Failure) it is impossible to discriminate $\ket{\phi^+}$ or $\ket{\phi^-}$ because all possible events of clicks from one can be also obtained from the other (double clicks at either upper or lower two detectors). (Loss) less than two clicks in all detectors indicates that photon loss occurs. Each detector is assumed to resolve up to two photons.

\begin{figure}
\centering
\includegraphics[width=3.2in]{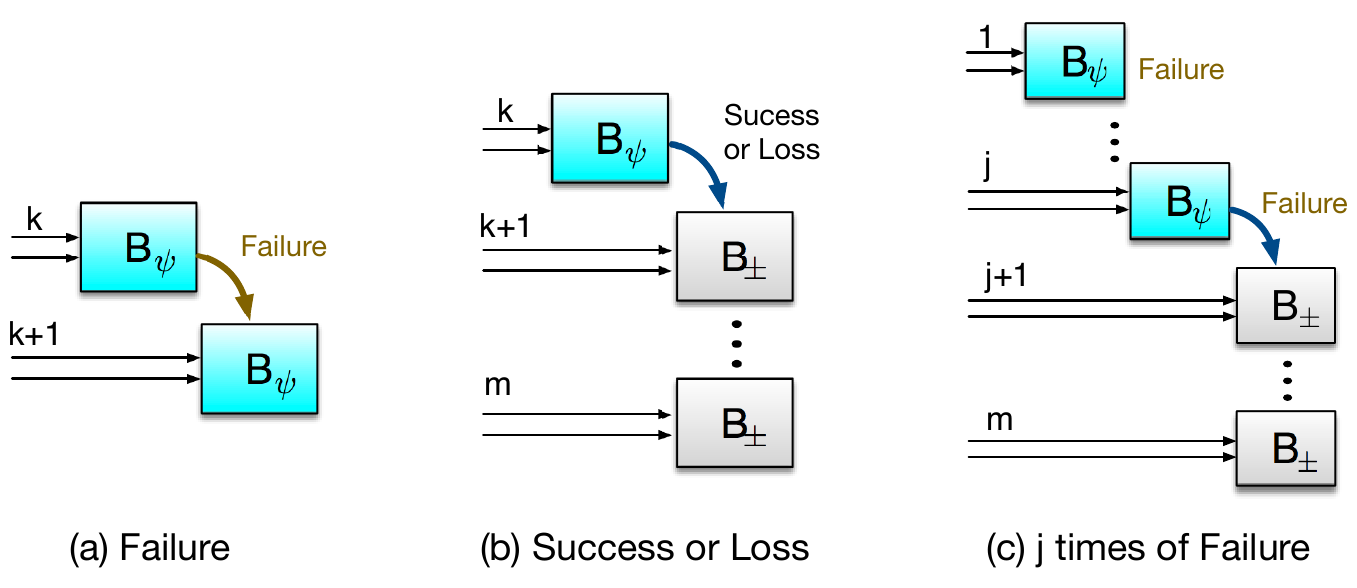}
\caption{In the 1st level Bell-state measurements ${\rm B}_{(1)}$, (a) if failure occurs at $k$-th ${\rm B}_{\psi}$ ($k=1,\dots,m$), we apply ${\rm B}_{\psi}$ again on next photon pair ($k+1$-th). (b) If $k$-th ${\rm B}_{\psi}$ succeeds with $\ket{\psi^{\pm}}$, apply ${\rm B}_{\pm}$ on all the remaining pairs (from $k+1$-th to $m$-th) together. If loss is detected at $k$-th ${\rm B}_{\psi}$, apply either ${\rm B}_{+}$ or ${\rm B}_{-}$ on all the remaining pairs together. (c) If ${\rm B}_{\psi}$ fail total $j$-times (from 1st to $j$-th), apply either ${\rm B}_{+}$ or ${\rm B}_{-}$ on all the remaining pairs together.
}\label{fig:aB1protocol}
\end{figure}

{\bf (1st level)} In the 1st level Bell measurement ${\rm B}_{(1)}$, $m$-times of ${\rm B}_{(0)}$ are performed on photon pairs, by following a simple rule illustrated in Fig.~\ref{fig:aB1protocol}: First, ${\rm B}_{\psi}$ is applied to arbitrary photon pair (one from the first qubit and the other from the second qubit). If $k$-th ${\rm B} _{\psi}$ ($k=1,\dots,m$) fails, ${\rm B}_{\psi}$ is applied again to $k+1$-th pair. If $k$-th ${\rm B}_{\psi}$ succeeds with $\ket{\psi^{\pm}}$, we apply ${\rm B}_{\pm}$ on all the remaining photon pairs (from $k+1$-th to $m$-th) together. If photon loss is detected at $k$-th ${\rm B}_{\psi}$, either ${\rm B}_{+}$ or ${\rm B}_{-}$ is selected arbitrarily and applied to all the remaining photon pairs (from $k+1$-th to $m$-th). If ${\rm B}_{\psi}$ fails consecutively $j$-times ($j$ is determined by optimization, $0 \leq j < m$), either ${\rm B}_{+}$ or ${\rm B}_{-}$ is arbitrarily selected and applied to all the remaining photon pairs (from $j+1$-th to $m$-th). By collecting all the results of $m$-times of ${\rm B}_{(0)}$, the result of ${\rm B}_{(1)}$ can be determined as follows: (i) Suceess - full discrimination of $\ket{\phi^{\pm}_{(m)}}$ and $\ket{\psi^{\pm}_{(m)}}$, (ii) Sign $\pm$ discrimination, or (iii) Failure.

{\bf (2nd level)} In the (logical) 2nd level Bell measurement ${\rm B}_{(2)}$, we perform $n$ independent $\rm B_{(1)}$. By collecting all their results, it is possible to identify the logical Bell states, $\ket{\Phi^{\pm}}$ and $\ket{\Psi^{\pm}}$.

\section{Success probabilities of CBM}
\label{asec:sPCBM}

We can calculate the success probabilities of CBM under the effects of photon losses.

\subsection{Without loss} 

We first assume that all the encoded photons are used in CBM without loss. In every $\rm B_{(1)}$, $\rm B_{\psi}$ is applied first total $j$ times. In this case, any single success of $\rm B_{\psi}$ can lead to unambiguous discrimination of $\ket{\phi_{(m)}^{\pm}}$ and $\ket{\psi_{(m)}^{\pm}}$: If any $\rm B_{\psi}$ succeeds with $\ket{\psi^{+}}$ ($\ket{\psi^{-}}$), one can find that the 1st level Bell state of this block is either $\ket{\phi_{(m)}^{+}}$ or $\ket{\psi_{(m)}^{+}}$ ($\ket{\phi_{(m)}^{-}}$ or $\ket{\psi_{(m)}^{-}}$) as shown in Eq.~(\ref{eq:level1Bellothers}). Then, by performing $\rm B_{+}$ ($\rm B_{-}$) on all the remaining photon pairs, it is possible to count the number of $\ket{\psi^{+}}$ ($\ket{\psi^{-}}$) contained in this block so that one can identify the result of $\rm B_{(1)}$ as
\begin{equation}
\nonumber
\begin{aligned}
&{\rm even~number~of~} \ket{\psi^{\pm}} \rightarrow \ket{\phi_{(m)}^{\pm}}\\
&{\rm odd~number~of~} \ket{\psi^{\pm}} \rightarrow \ket{\psi_{(m)}^{\pm}}.
\end{aligned}
\end{equation}
If all the $j$ times of $\rm B_{\psi}$ fail, we arbitrarily select and perform either $\rm B_{+}$ or $\rm B_{-}$ on the remaining photon pairs: If the selection is correct (with probability $1/2$), their results also lead to full discrimination of $\ket{\phi_{(m)}^{\pm}}$ and $\ket{\psi_{(m)}^{\pm}}$. Otherwise, only $\pm$ sign can be identified from the fact that a failure of $\rm B_{+}$ ($\rm B_{-}$) indicates that its sign is $-$ ($+$). Therefore, $\rm B_{(1)}$ discriminates $\ket{\phi_{(m)}^{\pm}}$ and $\ket{\psi_{(m)}^{\pm}}$ fully with probability $1-2^{-j-1}$, or identifies only the $\pm$ sign with probability $2^{-j-1}$. 

In the logical (2nd) level ${\rm B}_{(2)}$, total $n$ independent $\rm B_{(1)}$ are performed, whose result is either full or sign $\pm$ discrimination. Once any $\rm B_{(1)}$ yields success (full discrimination) with $\ket{\phi_{(m)}^{\pm}}$ or $\ket{\psi_{(m)}^{\pm}}$, one can discriminate between $\Phi$ and $\Psi$. Then, the sign $\pm$ can be identified by counting the total number of minus($-$) signs among the outcomes of $n$ times of $\rm B_{(1)}$ (see the Table~\ref{table:BMresult}): if even (odd) number of minus($-$) signs appear, the sign of logical Bell state is $+$($-$). For example, when $n=3$ and the outcomes of three $\rm B_{(1)}$ are $\{\ket{\phi_{(m)}^{-}},+,-\}$, the result of $\rm B_{(2)}$ is $\ket{\Phi^{+}}$ as $\phi$ and an even number of minus($-$) signs appears in the results. It fails only when all the $n$ independent $\rm B_{(1)}$ yield $\pm$ sign discriminations only with probability $2^{-(j+1)n}$. Therefore, if we set the protocol by $j=m-1$, the overall success probability of the logical Bell measurement is obtained as $P_s=1-2^{-nm}$.

\subsection{Under losses} 

Let us now consider CBM under photon losses. Assume that the photons in the first and second qubit survive with rate $\eta$ and $\eta'$, respectively. The success and failure probabilities of $\rm B_{(1)}$ can be calculated as below: 

In the assumption that any of the $2m$ photons (contained in two qubits) is not lost in each $\rm B_{(1)}$ with probability $(\eta\eta')^m$, full discrimination is possible as long as either any single $\rm B_{\psi}$ succeeds or $\rm B_{\pm}$ is chosen correctly (with $1/2$ probability) after $j$-times of failure of $\rm B_{\psi}$, so that the success probability can be written by
\begin{equation}
p_s(\eta,\eta')= \Big(1-\frac{1}{2^{j+1}}\Big)(\eta\eta')^m.
\label{eq:successprob}
\end{equation}

Note that photon loss in any ${\rm B}_{(0)}$ ($\rm B_{\psi}$ or $\rm B_{\pm}$) does not change the $\pm$ sign of the overall result of ${\rm B}_{(1)}$. Therefore, it turns out that the $\pm$ sign can be discriminated by either any single success of $\rm B_{\psi}$ or any single success or failure of $\rm B_{\pm}$ without loss. The failure of $\rm B_{(1)}$ occurs only in the case that all the performed $\rm B_{\psi}$ fail until loss is first detected and subsequently loss occurs in all $\rm B_{\pm}$ performed on the remaining photon pairs, so the failure probability can be directly written by
\begin{eqnarray}
p_f(\eta,\eta')=\sum^{m}_{l=m-j}\Big(\frac{1}{2}\Big)^{m-l}(\eta\eta')^{m-l}(1-\eta\eta')^l,
\label{eq:failprob1}
\end{eqnarray}
where $l$ indicates the number of $\rm B_{(0)}$ where photon loss occurs. An alternative way to calculate the failure probability is
\begin{equation}
\begin{aligned}
p_f(\eta,\eta')&=\mathop{\sum^{m}\sum^{m}}_{l_1+l_2=m-j}p(l_1,l_2)\binom{m}{l_1}\eta^{m-l_1}(1-\eta)^{l_1}\\
&\hspace{20mm}\times\binom{m}{l_2}\eta'^{m-l_2}(1-\eta')^{l_2},
\end{aligned}
\label{eq:failprob2}
\end{equation}
where $l_1$ and $l_2$ are the numbers of lost photons at first and second qubit, respectively. Note that there is no failure event for $l_1+l_2 \leq m-j-1$ and $\binom{a}{b}=0$ for $a<b$. Here, $p(l_1,l_2)$ can be calculated by counting all possible failure events as
\begin{equation}
\begin{aligned}
\nonumber
p(l_1,l_2)&=\sum^{{\rm min}[l_1+l_2,m]}_{l=m-j}\Big(\frac{1}{2}\Big)^{m-l}\binom{l}{l_d,l_1-l_d}/\binom{m}{l_1}\binom{m}{l_2}\\
&=\sum^{{\rm min}[l_1+l_2,m]}_{l=m-j}\Big(\frac{1}{2}\Big)^{m-l}\binom{l}{l_d}\binom{l-l_d}{l_1-l_d}/\binom{m}{l_1}\binom{m}{l_2}
\label{eq:failprobwithnum}
\end{aligned}
\end{equation}
where $l_d$ is the number of $\rm B_{(0)}$ where both photons are lost. One can easily verify that Eq.~(\ref{eq:failprob2}) is the same with the simple form in Eq.~(\ref{eq:failprob1}). The probability of only sign ($\pm$) discrimination can be obtained by $1-p_s(\eta,\eta')-p_f(\eta,\eta')$. 

In the logical (2nd) level, it is possible to discriminate $\ket{\Phi^{\pm}}$ and $\ket{\Psi^{\pm}}$ even under photon losses. Note that photon losses in any $\rm B_{(1)}$ does not affect the result of the other $\rm B_{(1)}$. Thus, by collecting the outcomes of the $n$ independent $\rm B_{(1)}$, one can fully discriminate $\ket{\Phi^{\pm}}$ and $\ket{\Psi^{\pm}}$ unless any $\rm B_{(1)}$ fails, or all the $n$ independent $\rm B_{(1)}$ yield $\pm$ sign discrimination only. Based on these, the overall success probability of CBM can be obtained by 
\begin{equation}
\label{eq:successP2}
P_s(\eta,\eta')=(1-p_f)^n-(1-p_s-p_f)^n,
\end{equation}
for given encoding parameters $(n, m, j)$ and transmission probabilities $\eta$ and $\eta'$. We can see that for $\eta=\eta'=1$ it becomes $P_s=1-2^{-nm}$ with $j=m-1$.

\subsection{Effect of logical errors}
\label{asec:LECBM}

General logical errors, bit or/and sign flips, may be produced in CBM due to experimental imperfections, depolarization, or operation errors. Any bit and sign flip errors in each physical (polarization) mode can induce also bit(symbol) flip($\phi \leftrightarrow \psi$) or/and sign ($+ \leftrightarrow -$) flip in the result of Bell measurement. Therefore, we need to carefully check and analyze the effects of general logical errors in CBM. We first assume that the bit and sign flip errors occur independently with rate $e_x$ and $e_z$ respectively in each photon mode. As an exemplary model, we can consider the depolarizing channel as $\rho \rightarrow (1-e_d)\rho+\frac{e_d}{4}\sum^3_{k=0}\sigma_k\rho\sigma_k$, where $\sigma_k \in \{\openone, X, Y, Z\}$ are Pauli operators. The independent bit and sign flip error rates in each photon mode can be then written by $e_x=e_d/2$ and $e_z=e_d/2$, respectively.

In the 0th level Bell measurement (i.e.~Bell measurement performed on photon pair, $\rm B_{(0)}$), both input photon modes may contain bit or/and sign flip errors, so that the errors are correlated in the result of Bell measurement. For example, if the first and second input photon modes experience $X\otimes Y$ ($Z\otimes Y$), logically sign (bit) flip would occur in the result of the Bell measurement. By taking into account all possible different error correlation of two input modes, we can obtain the logical error rates in $\rm B_{(0)}$ as $e^{(0)}_x=2e_x(1-e_x)$ and $e^{(0)}_z=2e_z(1-e_z)$.

In the 1st level $\rm B_{(1)}$, the bit flip error would propagate only through its success events, while in the case of sign $\pm$ discrimination and failure of $\rm B_{(1)}$ any bit flip of $\rm B_{(0)}$ does not affect the performance. The bit flip error occurs in  $\rm B_{(1)}$ if odd number of $\rm B_{(0)}$ contain bit flip errors, so that the rate is given by
\begin{equation} 
\nonumber
e_{x,s}^{(1)}=\sum_{p={\rm odd}}^{m}\binom{m}{p}(e_x^{(0)})^{p}(1-e_x^{(0)})^{m-p}.
\end{equation}
On the other hand, any sign $\pm$ flip errors in any $\rm B_{(0)}$ can be effectively corrected by majority vote among the success results of all $\rm B_{(0)}$ from the fact that the sign of all $\rm B_{(0)}$ should be the same in an ideal case due to our encoding strategy. The sign flip error propagates through success and sign discrimination results of $\rm B_{(1)}$ with different rates. The sign flip error rate, when $\rm B_{(1)}$ succeeds, can be obtained by counting all possible events that are not heralded by majority vote, as
\begin{equation}
\nonumber
\begin{aligned}
e_{z,s}^{(1)}&=\frac{1}{1-2^{-(j+1)}}\sum^j_{q=0}\frac{1}{2^{q+1}}\\
&\times\sum^{m-q}_{p=\lceil (m-q)/2 \rceil}\binom{m-q}{p}(e_z^{(0)})^{p}(1-e_z^{(0)})^{m-q-p}.
\end{aligned}
\end{equation}
When the result of $\rm B_{(1)}$ is sign discrimination, the effect of sign flip of $\rm B_{(0)}$ can be reduced by majority vote among the success outcomes of $m$-times of $\rm B_{(0)}$. The sign flip error rate in this case can be calculated by counting all possible events containing the unheralded errors as
\begin{equation}
\nonumber
\begin{aligned}
e_{z,\pm}^{(1)}&=\Big[\sum^{j-1}_{q=0}\Big(\frac{\eta\eta'}{2}\Big)^q\Big\{1-\frac{\eta\eta'}{2}-\frac{(\eta\eta')^{m-q}}{2}-(1-\eta\eta')^{m-q}\Big\}\\
&\times\sum^{m-1-q}_{p=\lceil \frac{m-1-q}{2} \rceil}\binom{m-1-q}{p}(e_z^{(0)})^{p}(1-e_z^{(0)})^{m-1-q-p}\\
&+\Big(\frac{\eta\eta'}{2}\Big)^j\Big\{1-\frac{(\eta\eta')^{m-j}}{2}-(1-\eta\eta')^{m-j}\Big\}\\
&\times\sum^{m-j}_{p=\lceil \frac{m-j}{2} \rceil}\binom{m-j}{p}(e_z^{(0)})^{p}(1-e_z^{(0)})^{m-j-p}\Big]\\
&\hspace{10mm}/(1-p_s-p_f).
\end{aligned}
\end{equation}

In the 2nd (logical) level, bit flip errors in the results of $\rm B_{(1)}$ can be heralded by majority vote among the success outcomes due to the fact that the symbol either $\Phi$ or $\Psi$ of all $\rm B_{(1)}$ should be the same in our encoding strategy. The sign flip errors of $\rm B_{(1)}$ would induce also sign flip in the result of CBM if odd number of $\rm B_{(1)}$ contain sign flip errors. Two error rates can be then calculated by
\begin{equation}
\nonumber
\begin{aligned}
e_{s,x}^{(2)}&=\frac{1}{P_s(\eta,\eta')}\sum^n_{k=1}\binom{n}{k}p^k_s(1-p_s-p_f)^{n-k}Xr(k)\\
e_{s,z}^{(2)}&=\frac{1}{P_s(\eta,\eta')}\sum^n_{k=1}\binom{n}{k}p^k_s(1-p_s-p_f)^{n-k}Zr(k),
\end{aligned}
\end{equation}
where 
\begin{equation}
\nonumber
\begin{aligned}
Xr(k)&=\sum^k_{l=\lceil k/2 \rceil}\binom{k}{l}(e_{x,s}^{(1)})^{l}(1-e_{x,s}^{(1)})^{k-l}\\
Zr(n,k)&=\sum^k_{p={\rm odd\geq1}}\binom{k}{p}(e_{z,s}^{(1)})^{p}(1-e_{z,s}^{(1)})^{k-p}\\
&\times\sum_{q={\rm even\geq2}}^{n-k}\binom{n-k}{q}(e_{z,\pm}^{(1)})^{q}(1-e_{z,\pm}^{(1)})^{n-k-q},\\
&+\sum^k_{p={\rm even\geq2}}\binom{k}{p}(e_{z,s}^{(1)})^{p}(1-e_{z,s}^{(1)})^{k-p}\\
&\times\sum_{q={\rm odd\geq1}}^{n-k}\binom{n-k}{q}(e_{z,\pm}^{(1)})^{q}(1-e_{z,\pm}^{(1)})^{n-k-q}.
\end{aligned}
\end{equation}
The overall success probability of CBM can be divided into the probabilities that contain each Pauli logical errors as $P_s(\eta,\eta')=P_{s,i}+P_{s,x}+P_{s,y}+P_{s,z}$ where 
\begin{equation}
\label{eq:Perror}
\begin{aligned}
P_{s,x}&=\sum^n_{k=1}\binom{n}{k}p^k_s(1-p_s-p_f)^{n-k}Xr(k)\big(1-Zr(n,k)\big)\\
P_{s,y}&=\sum^n_{k=1}\binom{n}{k}p^k_s(1-p_s-p_f)^{n-k}\big(1-Xr(k)\big)Zr(n,k)\\
P_{s,z}&=\sum^n_{k=1}\binom{n}{k}p^k_s(1-p_s-p_f)^{n-k}Xr(k)Zr(n,k)\\
P_{s,i}&=\sum^n_{k=1}\binom{n}{k}p^k_s(1-p_s-p_f)^{n-k}\big(1-Xr(k)\big)\big(1-Zr(n,k)\big).
\end{aligned}
\end{equation}
As a result, in principle, any possible flip and sign errors in lower level (0th or 1st) can be reduced in the (logical) 2nd level by majority vote among the results of lower level Bell measurements (as long as $n\geq3$). The errors can be corrected more effectively if increasing the encoding size $(n,m)$. It shows a tendency that bit flip errors are reduced further when $j$ parameter increases for a given $m$, as it yields more success events of $\rm B_{(1)}$ in the logical  level. The performance would depend on the given error rates, $e_d$ in polarizing channels and $\eta$ and $\eta'$ of two qubits under losses. The optimal strategy with $(n,m,j)$ would thus differ according to the purpose of the applications. 

\subsection{Effect of dark counts} 
\label{asec:dcCBM}

\begin{figure}
\centering
\includegraphics[width=2.6in]{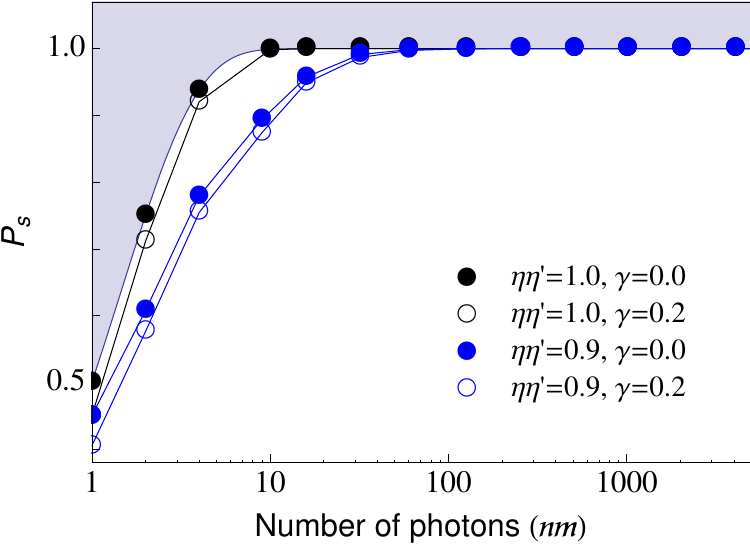}
\caption{Effect of dark counts on the success probability $P_s$ under loss $\eta\eta'$.
}\label{fig:FigDC}
\end{figure}

Dark count is the click when no photon is present, a possible imperfection of photodetectors. Note that there is no additional input of photons (no ancillary input) during the process in which the number of photons is restricted. Moreover, if the logical Bell states are generated with photon pairs from a typical down-conversion, the probability to contain more than two photons in a single mode is negligible and can be heralded in the generation process. Dark count is typically of thermal origin. We assume that the dark count rate at each photodetector $\lambda$ is enough small so that the probability that dark counts occur at more than two detectors simultaneously among four detectors in each $\rm B_{(0)}$ is negligible, i.e. $6\lambda^2(1-\lambda)^2+4\lambda^3(1-\lambda)+\lambda^4\sim 0$. We thus define again the overall dark count rate of each $\rm B_{(0)}$ as $\gamma\equiv4\lambda(1-\lambda)^3$.

In $\rm B_{(0)}$, any separated clicks of two photons i.e.~(H,V), (V,H), (H,H), (V,V) - one click from the upper two detectors and another from the lower two - are the success events, while double clicks at either upper or lower detectors e.g.~(HV,0) or (0,2V) are failure. If an additional click occurs at one of the detectors, the results can be changed: Half of the original success events are changed to failure by an additional click e.g. (HV, H), (HV,V), (H, HV), (V,HV), while the other half events remain as success e.g. (2H,V), (H,2V) regarded as (H,V) success. Original failure events plus a dark count yield 75\% failure e.g.~(HV,V), (0,H2V) and 25\% success e.g.~(2H,V), (H,2V). In this case, the success event would produce either sign or bit flip errors with probability $1/2$. A dark count compensating a photon loss yields normal two clicks, 50\% success and 50\% failure. In this case, the success would produce also logical errors (either sign or bit flip) with probability $1/2$ (we will handle such a logical error correction in the following subsection). As a result, the original success and failure probabilities of $\rm B_{(0)}$, $1/2$, are changed respectively to
\begin{equation}
\nonumber
\begin{aligned}
&\Big\{\frac{1}{2}(1-\gamma)+(\frac{1}{2}\times\frac{1}{4}+\frac{1}{2}\times\frac{1}{2})\gamma\Big\}\eta\eta'\\
&+\frac{1}{2}\gamma\{\eta(1-\eta')+(1-\eta)\eta'\}\sim\Big(\frac{1}{2}-\frac{\gamma}{8}\Big)\eta\eta',\\
&\Big\{\frac{1}{2}(1-\gamma)+(\frac{1}{2}\times\frac{3}{4}+\frac{1}{2}\times\frac{1}{2})\gamma\Big\}\eta\eta'\\
&+\frac{1}{2}\gamma\{\eta(1-\eta')+(1-\eta)\eta'\}\sim\Big(\frac{1}{2}+\frac{\gamma}{8}\Big)\eta\eta',
\end{aligned}
\end{equation}
with transmission rates $\eta$ and $\eta'$ of two qubits, where the events of compensation by losses, $\gamma(1-\eta)$ and $\gamma(1-\eta')$, are negligible. Now, we can calculate the success probability of $\rm B_{(1)}$ as
\begin{equation}
\nonumber
\begin{aligned}
p_s(\eta,\eta';\gamma)&\sim \Big[1-\Big(\frac{1}{2}+\frac{\gamma}{8}\Big)^{j}\Big\{\frac{1}{2}+\frac{1}{2}\sum^{m-j}_{k=1}\Big(\frac{1}{2}\Big)^k\binom{m-j}{k}\\
&\times\gamma^k(1-\gamma)^{m-j-k}\Big\}\Big]\eta^m\eta'^m,
\label{eq:successprobdarkcount}
\end{aligned}
\end{equation}
where the second term in $\{\cdot\}$ is due to the additional failure induced by dark counts in each $\rm B_{\pm}$. It becomes equivalent with Eq.~(\ref{eq:successprob}) when $\gamma=0$. Similarly we can calculate the failure probability of $\rm B_{(1)}$ as
\begin{eqnarray}
\nonumber
p_f(\eta,\eta')=\sum^{m}_{l=m-j}\Big(\frac{1}{2}+\frac{\gamma}{8}\Big)^{m-l}(\eta\eta')^{m-l}(1-\eta\eta')^l.
\label{eq:failprob1dark}
\end{eqnarray}
The overall success probability of CBM under dark counts and losses is plotted in Fig.~\ref{fig:FigDC}. In shows that the effect of dark counts on the performance of CBM is small and diminished further as the encoding size increases.

\section{Building blocks for quantum network}

\subsection{Extending communication range by CBM}
\label{asec:extension}

\begin{figure}
\centering
\includegraphics[width=2.8in]{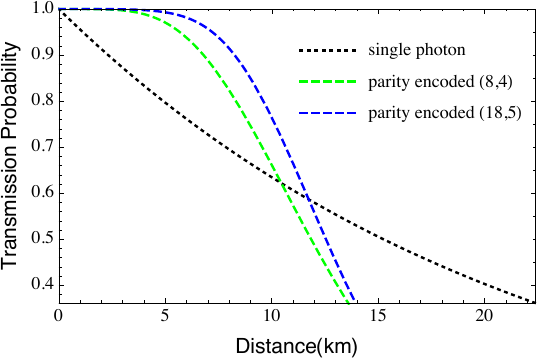}
\caption{Direct transmission of parity encoded qubits. The parity encoded qubits can travel with higher transmission probabilities (rates) than single photon transmission within some limited distance range depending on the encoding size.
}\label{fig:ComparisionParitySingle}
\end{figure}

\begin{figure}
\centering
\includegraphics[width=3.2in]{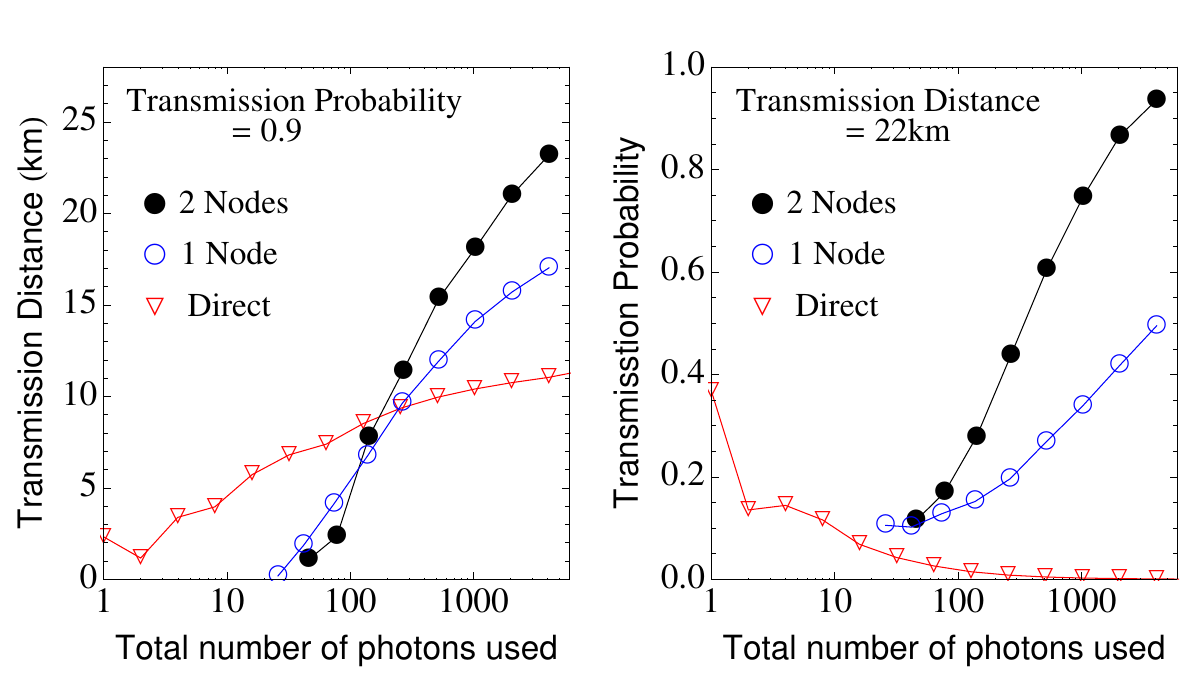}
\caption{Enhancement of the transmission probability and distance by CBM. The performances between the direct transmission of parity encoded qubits and the transmission assisted by CBM with equally separated nodes. Left: The maximal transmission distances with probability 0.9 is plotted against the total number of photons used in the process. Right: The maximal transmission probabilities to transmit to 22 km is plotted. We here assumed $\eta_0=1$.
}\label{fig:ComparisionTrans}
\end{figure}

Let us check if CBM is useful to extend the communication range over the direct transmission. A single photon can travel to distance $L$ with probability $e^{-L/L_{\rm att}}$, which decays exponentially over distance. A parity encoded photonic qubit can travel to distance $L$ if two requirements are met i) at least one photon arrives in each block, and ii) at least one block arrives without loss, so that the transmission probability is $P_{\rm direct}=(1-(1-\eta)^m)^n-(1-(1-\eta)^m-\eta^m)^n$ where $\eta=e^{-L/L_{\rm att}}$. A significantly higher transmission probability than single photon transmission can be achieved within some limited distance range as compared in Fig.~\ref{fig:ComparisionParitySingle}. If CBM is applied at intermediate  nodes (also at the final location), the transmission probability is changed to $P_s(\eta_{L_0},\eta_0)^{d+1}$ where $d=L/L_0-1$ is the number of nodes. In Fig.~\ref{fig:ComparisionTrans}, we compare the direct transmission and the transmission assisted by CBM at intermediate nodes ($d=1,2$), regarding the maximal transmission distances and probabilities with a fixed number of available photons in total. It clearly shows that the scheme assisted by CBM can enhance the communication range and rate over direct transmissions.

\subsection{Estimated time($\tau_p$) to generate $\ket{\Phi^+}$}
\label{asec:timetau}

Assuming that $\ket{\Phi^+}$ is generated from photon pairs, we can estimate the taken time $\tau_p$:

i) Suppose that entangled photon pairs in the polarization degree of freedom are prepared through a typical down-conversion scheme. By applying Type-I fusion gate based on linear optics (defined in Ref.~\cite{DEBrown2005}) on photon pairs, GHZ states with $m+1$ photons (the coefficient $1/\sqrt{2}$ will be omitted hereafter), $\ket{{\rm GHZ}_{m+1}}=\ket{H}^{\otimes m}\ket{+} +\ket{V}^{\otimes m}\ket{-}$, can be generated, in which a single redundant mode is prepared in basis $\ket{\pm}=\ket{H}+\ket{V}$. Note that total $m-1$ times of fusion gate operations are applied on photon pairs to generate $\ket{{\rm GHZ}_{m+1}}$.

ii) Then, two $\ket{{\rm GHZ}_{m+1}}$ states can be merged by applying Type-I gate on their redundant modes, resulting in
\begin{equation}
\begin{aligned}
&(\ket{H}^{\otimes m}+\ket{V}^{\otimes m})(\ket{H}^{\otimes m}+\ket{V}^{\otimes m})\ket{H}\\
&+(\ket{H}^{\otimes m}-\ket{V}^{\otimes m})(\ket{H}^{\otimes m}-\ket{V}^{\otimes m})\ket{V},
\label{eq:middletoparity}
\end{aligned}
\end{equation}
composed of two blocks of $m$-photon GHZ states and a redundant mode.

iii) Likewise, by applying Type-I gates on redundant modes of states prepared in (\ref{eq:middletoparity}), one can produce
\begin{equation}
(\ket{H}^{\otimes m}+\ket{V}^{\otimes m})^{\otimes 2n}\ket{H}+(\ket{H}^{\otimes m}-\ket{V}^{\otimes m})^{\otimes 2n}\ket{V},
\label{eq:paritywithredun}
\end{equation}
with total $n-1$ times of gate operations. 

iv) Finally, by removing the redundant mode with a single photon measurement, the logical Bell pair $\ket{\Phi^+}$ is obtained. In the assumption that we perform the fusion gate operations in parallel based on the knock-down tournament type procedure \cite{Vernava2007}, the average total time taken to generate $\ket{\Phi^+}$ can be estimated as
 \begin{equation}
 \begin{aligned}
\tau_p(n,m)&\simeq\Big(\lceil \log_2{m} \rceil+1+\lceil \log_2{n} \rceil+1\Big)\tau\\
&=\Big(\lceil \log_2{m} \rceil+\lceil \log_2{n}\rceil+2\Big)\tau,
\label{eq:generationtime}
\end{aligned}
\end{equation}
where each terms in the first equation indicates the time cost for each steps from i) to iv), respectively.

\subsection{Numerical optimization}
\label{asec:Nume}

The total cost of photons for transmiting a logical qubit over $L$ can be estimated in average \cite{Munro12,Azuma15} as
\begin{equation}
Q=\frac{2nm}{Rt_0}\times\frac{L}{L_0}.
\end{equation}
where $2nm$ is the number of photons in a logical Bell pair and $L/L_0$ is the number of intermediate nodes plus a sender node, and $Rt_0$ is the overall success probability. Note that it does not contain the photons consumed to prepare the Bell pairs in each nodes.

We perform numerical searches to optimize our protocol to transmit a qubit over distance $L$. The optimized parameters $\{n,m,j,L_0\}$ are determined to minimize the total cost of photons $Q$, by taking into account possible losses and errors on both qubits during transmission and operations in the repeater. The minimized $Q$ is evaluated by
\begin{equation}
Q_{\rm min}\equiv \underset{n,m,j,L_0}{\rm min}Q(n,m,j,L_0).
\end{equation}
by numerical searches over $\{n,m,j,L_0\}$. For the general logical errors, we model the error in each photon mode by a typical depolarizing channel $\rho \rightarrow (1-e_d)\rho+\frac{e_d}{4}\sum^3_{k=0}\sigma_k\rho\sigma_k$, where $\sigma_k$ are Pauli operators. The error rates are $e_x=e_d/2$ and $e_z=e_d/2$ in each physical photon mode, and $e^{(0)}_x=e_d(1-e_d/2)$ and $e^{(0)}_x=e_d(1-e_d/2)$ in the 0th level of CBM ($\rm B_{(0)}$). Given depolarizing error rate $e_d$, we can obtain the average fidelity $F$ of the transmission over distance L (following the analyses in Appendixes~\ref{asec:LECBM} and \ref{asec:elogical}). 

In numerical searches, we set some parameters as below: source and detector inefficiencies $\epsilon_s\epsilon_d$=1 or 0.95, depolarizing error rate $e_d=5.6\times10^{-5}$, the time taken in measurements $\tau=150$ ns, the attenuation length $L_{\rm att}=22$ km, and the speed of light in optical fiber $c=2\times 10^8 {\rm ms}^{-1}$. We consider different examples of transmission distances and obtain the results as follows:

i) For 1,000 km, 
\begin{itemize}
\item[-] $Q_{\rm min}$=$1.3\times10^5$, $Rt_0$=0.702, $F=0.98$, $L_0$=1.7 km, $n$=13, $m$=6, $j$=2, $\tau_p$=1.35$\mu$s, $\epsilon_s \epsilon_d$=1.0, $\eta_0=0.986$
\item[-] $Q_{\rm min}$=$7.4\times10^5$, $Rt_0$=0.700, $F=0.96$, $L_0$=1.8 km, $n$=58, $m$=8, $j$=1, $\tau_p$=1.65$\mu$s, $\epsilon_s \epsilon_d$=0.95, $\eta_0=0.934$
\end{itemize}

ii) For 5,000 km,
\begin{itemize}
\item[-] $Q_{\rm min}$=$1.0\times10^6$, $Rt_0$=0.798, $F=0.97$, $L_0$=1.4 km, $n$=16, $m$=7, $j$=2, $\tau_p$=1.35$\mu$s, $\epsilon_s \epsilon_d$=1.0, $\eta_0=0.986$
\item[-] $Q_{\rm min}$=$7.4\times10^6$, $Rt_0$=0.669, $F=0.93$, $L_0$=1.5 km, $n$=83, $m$=9, $j$=1, $\tau_p$=1.95$\mu$s, $\epsilon_s \epsilon_d$=0.95, $\eta_0=0.932$
\end{itemize}

iii) For 10,000 km,
\begin{itemize}
\item[-] $Q_{\rm min}$=$2.4\times10^6$, $Rt_0$=0.773, $F=0.97$, $L_0$=1.2 km, $n$=16, $m$=7, $j$=2, $\tau_p$=1.35$\mu$s, $\epsilon_s \epsilon_d$=1.0, $\eta_0=0.986$
\item[-] $Q_{\rm min}$=$1.9\times10^7$, $Rt_0$=0.698, $F=0.92$, $L_0$=1.4 km, $n$=92, $m$=10, $j$=2, $\tau_p$=1.95$\mu$s, $\epsilon_s \epsilon_d$=0.95, $\eta_0=0.932$
\end{itemize}

\subsection{Transmission fidelity}
\label{asec:elogical}

We also need to take into account the effects of logical errors on the performance. The errors in each repeater nodes propagate along the network to Bob, so the total success probability can be divided into the transmission probabilities with each Pauli errors by $P^{\rm tot}_s=P^{\rm tot}_{s,i}+P^{\rm tot}_{s,x}+P^{\rm tot}_{s,y}+P^{\rm tot}_{s,z}=(P_{s,i}+P_{s,x}+P_{s,y}+P_{s,z})^{L/L_0}$, where we use $P_s(\eta_{L_0},\eta_0)=P_{s,i}+P_{s,x}+P_{s,y}+P_{s,z}$ obtained in Appendix~\ref{asec:LECBM}. From the fact that each logical error at receiver (Bob) occurs when total odd number of repeater nodes produce the logical errors, one can calculate the effective rate of bit and sign flip errors divided by the overall transmission probability as
\begin{equation}
Q_{X/Z}=\frac{1}{2}\Big[1-\frac{(P_{s,i}\mp P_{s,x}\pm P_{s,z}-P_{s,y})^{L/L_0}}{(P_{s,i}+P_{s,x}+P_{s,y}+P_{s,z})^{L/L_0}}\Big].
\end{equation} 
The overall transmission process with logical errors can be then modeled by ${\cal E}(\rho)=P_{I}\rho+P_{X}X\rho X+P_{Y}Y\rho Y+P_{Z}Z\rho Z $ with probabilities $P_{I}=P^{\rm tot}_{s,i}/P^{\rm tot}_s=(1-Q_{X})(1-Q_{Z})$, $P_{X}=P^{\rm tot}_{s,x}/P^{\rm tot}_s=Q_{X}(1-Q_{Z})$, $P_{Y}=P^{\rm tot}_{s,y}/P^{\rm tot}_s=(1-Q_{X})Q_{Z}$, $P_{Z}=P^{\rm tot}_{s,z}/P^{\rm tot}_s=Q_{X}Q_{Z}$.
The average fidelity of the transmission is then given as $F=P_{I}$. For the application to quantum key distribution in our protocol, the key generation rate can be asymptotically obtained by $R= {\rm max}[P^{\rm tot}_s\big\{1-2h(Q)\big\}/t_0,0]$ with the binary entropy function $h(Q)=-Q\log_2(Q)-(1-Q)\log_2(1-Q)$ and $Q=(Q_X+Q_Z)/2$.

\section{Comparison with other recent proposals}
\label{asec:comp}

\subsection{Loss-tolerant Bell measurements}
\label{asec:compBM}

\begin{table*}[t]
\caption{\label{table:comparesp}The success probabilities achieved with encoding $(n,m)$ under photon losses.}
\begin{ruledtabular}
\begin{tabular}{rrlrlrlrlrl}
&\multicolumn{2}{c}{$\eta=1$}&\multicolumn{2}{c}{$\eta=0.99$}&\multicolumn{2}{c}{$\eta=0.95$}&\multicolumn{2}{c}{$\eta=0.9$}&\multicolumn{2}{c}{$\eta=0.75$}\\
$(n,m)$&CBM&\cite{Ewert16}\footnote{a scheme proposed by Ewert {\em et al.} in Ref.~\cite{Ewert16}}
&CBM&\cite{Ewert16}&CBM&\cite{Ewert16}&CBM&\cite{Ewert16}&CBM&\cite{Ewert16}\\
\hline
(1,1) &50&(50)&49.5&(49.5)&47.5&(47.5)&45&(45)&37.5&(37.5)\\
(2,2)&93.75&(75)&92.24&(73.99)&86.01&(69.66)&77.91&(63.79)&53.39&(44.82)\\
(3,10)&100.00&(87.5)&99.91&(83.56)&93.49&(65.61)&72.31&(43.71)&15.94&(8.21)\\
(6,5)&100.00&(98.44)&100.00&(97.91)&99.87&(94.69)&98.57&(87.74)&74.56&(52.86)\\
(10,3)&100.00&(99.90)&99.95&(99.87)&99.51&(99.51)&97.95&(97.95)&77.77&(77.77)\\
(23,5)&100.00&(100.00)&100.00&(100.00)&100.00&(100.00)&99.95&(99.95)&93.50&(92.44)
\end{tabular}
\end{ruledtabular}
\end{table*}

Advanced Bell measurement schemes have been proposed recently to achieve high success probabilities beyond 50\% limit based on linear optics. The highest probability achieved so far is $1-2^{-N}$, with $N$-photon entanglement \cite{SLee15}. In fact, (as we proved in this article) the success probability $1-2^{-N}$ is the fundamental upper bound limited by linear optics with $N$-photon encoding. However, when photon loss occurs, the scheme in \cite{SLee15} requires an additional encoding for error correction. A scheme proposed by Ewert {\em et al.}~\cite{Ewert16} similarly employs a multiphoton entanglement based on the fact that a parity encoded qubit is robust to losses \cite{Ralph2005}. It achieves the success probability up to $1-2^{-n}$ with $N=nm$ photons per qubit, and tolerates some losses by means of redundantly encoded photons. In this scheme, however, among total $N=nm$ photons in a logical qubit, at best $n$ photons (one from each $n$ blocks) can contribute to enhance the success probability, while the others are consumed redundantly.

CBM allows us to reach the upper bound of the success probability by linear optics, $1-2^{-N}$, when total $N=nm$ photons are used per qubit. Arbitrary high success probabilities up to unit can be achieved even under photon losses by increasing $N$ without additional error correction, as long as it satisfies the no-cloning theorem, i.e. $\eta\eta'>0.5$. To our knowledge, CBM is so far the only Bell measurement that enables to saturate both fundamental limits by linear optics and no-cloning theorem. From a practical point of view, CBM outperforms all the other proposals with respect to the achieved success probability using the same number of photons in total and under the same loss rate. In Table~\ref{table:comparesp}, the success probabilities of optimized CBM and the scheme in \cite{Ewert16} are compared for the same encoding size $(n,m)$ and loss rate $\eta$. We note that in CBM photons effectively contribute to either increase the success probability or protect the qubit from losses, in contrast to the scheme in \cite{Ewert16}, which consumes at least $n(m-1)$ photons redundantly. 

\subsection{3rd generation quantum repeaters}
\label{asec:compR}

\begin{figure}[b]
\centering
\includegraphics[width=3.6in]{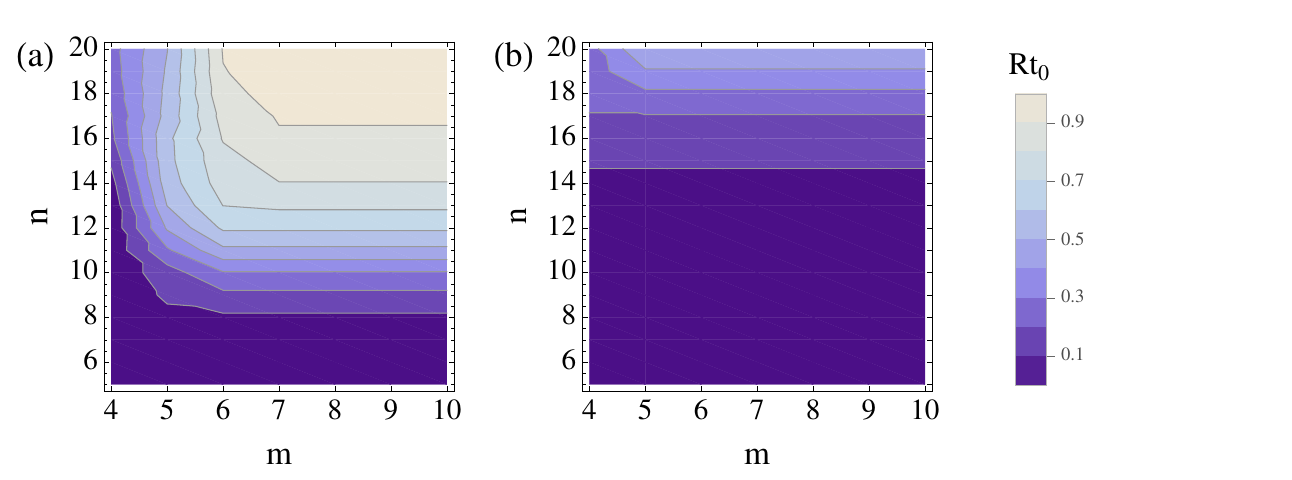}
\caption{Maximum transmission probabilities $Rt_0$ over 1,000 km using quantum repeater ($L_0=1.7$) based on (a) CBM and (b) the scheme proposed in \cite{Ewert16}. We set $\eta_0=0.99$ on both qubits, by taking into account the losses and imperfections during the preparation and measurement processes in each repeater.
}\label{fig:comp1000}
\end{figure}

Various quantum repeater protocols \cite{Munro12,Muralidharan14,Azuma15,Ewert16} have been proposed recently based on quantum error corrections and multi-photon encoding that can be used, in principle, to correct losses and errors at each repeater stations. In contrast to the standard quantum repeater protocols \cite{DLCZ,Briegle98,Sangouard11}, quantum repeaters developed in this direction, sometimes refereed as {\em 3rd generation quantum repeaters} \cite{Muralidharan16}, do not necessitate round trip heralding signals between nodes and long-lived quantum memories. It may be thus expected that such a repeater, as the speed of communication is (in principle) limited by only the local operation time, will be able to considerably enhance the performance of quantum communication within polynomial scaling over distances.

All-optical quantum repeaters, categorized also as 3rd generation, have been proposed recently \cite{Azuma15,Ewert16} and some preliminary models are experimentally demonstrated \cite{Yasushi19,Zehng19}. A repeater protocol based on optical systems provides some advantages as discussed in \cite{Azuma15,Ewert16}: it can be performed by photon sources, linear optical elements, and photon detectors. Since a deterministic conversion between photon and matter qubits are demanding, all-optical operations with only photonic qubits at room temperature may be quite an attractive route to quantum repeaters compared to matter-based approaches. However, besides the requirement of efficient generation of photon resources, there exist two major difficulties to overcome in all-optical approaches: (i) at best 50\% success probability of the Bell measurement on single photons, and (ii) photon losses not only during transmissions between repeaters but also during the stationary process in the repeater. The protocol proposed by Azuma {\em et al.} \cite{Azuma15} takes a time-reversal approach with the help of photonic cluster states to avoid the probabilistic nature of Bell measurement. Against photon losses, additional loss-tolerant encoding and feedforward tactics proposed by Vernava {\em et al.} \cite{Vernava2007} are employed. It could achieve comparable communication rates and resource costs with the speediest matter-based protocol by Munro {\em et al.} \cite{Munro12}. The proposal by Ewert {\em et al.} \cite{Ewert16} is based on an advanced scheme of Bell measurement and the parity state encoding (similar to ours as we showed in the previous section). The repeater is designed for one-way quantum communication along the network, in which quantum teleportation is performed with the Bell measurement on the arriving encoded qubit and one qubit from the prepared entangled encoded pair. It is claimed that, in principle, an ultra-fast communication is possible without feedforward assuming instant generations of entangled photons.
Although the abovementioned proposals provide advanced protocols to achieve considerably fast and resource-efficient communications over the conventional quantum repeaters, neither of them address the ultimate limits of the performance in designing quantum network following this route. 

Our Bell measurement scheme (CBM) can be directly used as a building block for all-optical quantum network, either for the entanglement distribution between Alice and Bob or for one-way transmission of a qubit across the network. Either designs have exactly the same success probabilities and performances in our protocol, so that any type can be chosen depending on the purpose of the applications. This is in contrast to other proposals; the protocol by Azuma {\em et al.} seems to be more suitable for entanglement distribution, and the protocol by Ewert {\em et al.} is designed for one-way transmission. Moreover, since CBM could reach both fundamental upper bounds (efficiency and loss-tolerance), its performance in the application to quantum repeaters would exceed other proposals. For analyzing the performance, in contrast to the analysis based other protocols \cite{Ewert17}, we take into account all possible errors, imperfections and losses on both qubits, not only during the transmission but also during the process in the repeaters. This may be reasonable that, in all-optical implementations, losses may be quite detrimental even to the qubit staying in the repeater (i.e.,~stationary qubit) for the entanglement generation and measurement process. In this circumstance, we first compare the maximum transmission probabilities over 1,000 km between ours and the proposal in \cite{Ewert16}, obtained by numerical searches for a given encoding size $(n,m)$ and with 1\% overall loss and imperfection on both qubits in the repeater ($\eta_0=0.99$), as shown in Fig.~\ref{fig:comp1000}. It shows that ours could achieve almost near-deterministic communication probabilities within moderate encoding sizes, while the protocol by \cite{Ewert16} reaches around 50\% probability with the same encoding size. This indicates that our protocol performs better in more realistic models. We then fully analyze and compare the performance for arbitrary long distance quantum communications in terms of the optimal strategy to minimize the photon cost overall. For fair comparison, we take the parameters which are the same as or properly selected from the proposals in \cite{Azuma15}. Our optimal protocol for the communication length $L=$ 5,000 (1,000) km yields $Rt_0=$ 0.70 (0.70) with total $Q_{\rm min}= 7.4 \times 10^6$ ($7.4 \times 10^5$) number of photons. If we compare these with the results in \cite{Azuma15}, $Q_{\rm min}= 4.0 \times 10^7$ ($4.1 \times 10^6$) with $Rt_0=$ 0.69 (0.58), our protocol costs only one order of magnitude less photons ($\sim18\%$ photons) to achieve comparable transmission probabilities with the optimal protocol in \cite{Azuma15}. If the same order of photon are used, much higher (up to unit) transmission probability can be attained with our scheme. The transmission rate $R$ is dependent on the local operation time as $\sim1/t_0$ in the repeater. The required components in our repeater is photon source, linear optics, and photon detection (with one or two-step feedforward of wave plate modulation). As CBM plays a role both for the logical Bell measurement and error corrections, neither of additional gate operations nor quantum memories are necessary in our protocol.




\end{document}